\documentclass[sigconf]{acmart}

\usepackage{xr}
\usepackage{hyperref}

\usepackage{tabu}                      
\usepackage{booktabs}                  
\usepackage{xspace}

\newcommand{\drag}{Drag--\&--Flick\xspace}
\newcommand{\push}{Push--\&--Release\xspace}
\newcommand{\head}{Head-Polynomial\xspace}

\usepackage[normalem]{ulem}
\usepackage{amsmath}

\newcommand{\etodo}[1]{}

\definecolor{deepgreen}{RGB}{0,150,0}  

\newcommand{\update}[1]{\textcolor{black}{#1}}
\newcommand{\updateNEW}[1]{\textcolor{black}{#1}}

\newcommand{\degree}{\textbf{$^{\circ}$~}}

\AtBeginDocument{%
  \providecommand\BibTeX{{%
    \normalfont B\kern-0.5em{\scshape i\kern-0.25em b}\kern-0.8em\TeX}}}

\copyrightyear{2026}
\acmYear{2026}
\setcopyright{cc}
\setcctype{by}
\acmConference[CHI '26]{Proceedings of the 2026 CHI Conference on Human Factors in Computing Systems}{April 13--17, 2026}{Barcelona, Spain}
\acmBooktitle{Proceedings of the 2026 CHI Conference on Human Factors in Computing Systems (CHI '26), April 13--17, 2026, Barcelona, Spain}
\acmPrice{}
\acmDOI{10.1145/3772318.3791549}
\acmISBN{979-8-4007-2278-3/2026/04}

\begin{document}


\title{Leveraging Head Movement for Navigating Off-Screen Content on Large Curved Displays}
\renewcommand{\shorttitle}{Leveraging Head Movement for Navigating Off-Screen Content on Large Curved Displays}

\author{A K M Amanat Ullah}
\affiliation{%
  \institution{The University of British Columbia}
  \streetaddress{}
  \city{British Columbia}
  \country{Canada}}
\email{amanat7@student.ubc.ca}
\orcid{0000-0001-5402-0160}

\author{David Ahlström}
\affiliation{%
  \institution{University of Klagenfurt}
  \streetaddress{}
  \city{Klagenfurt}
  \country{Austria}}
\email{david.ahlstroem@aau.at}
\orcid{0000-0002-9553-1685}

\author{Khalad Hasan}
\affiliation{%
  \institution{The University of British Columbia}
  \city{British Columbia}
  \country{Canada}}
\email{khalad.hasan@ubc.ca}
\orcid{0000-0002-4815-5461}

\renewcommand{\shortauthors}{Ullah, Ahlström and Hasan}

\begin{abstract}

Large curved displays are ideal for viewing 360\degree content, such as 3D maps, but typically restrict users to a 180\degree viewport, leaving information off-screen. Since users naturally direct their heads toward regions on-screen before interacting, head movements offer a promising alternative for workspace manipulation to bring off-screen content into view. We explore rate control functions (linear, sigmoid, polynomial) and zone control functions (continuous, friction, interrupted, additive) to translate head rotations into workspace control, enabling users to access off-screen content. Polynomial rate control emerges as the best choice, achieving the fastest trial times and highest subjective ratings. Using a map navigation task, our second study demonstrates that users perform better with the polynomial head-based technique than with the industry-standard controller-based methods, click-and-drag and joystick-push, for 360\degree workspace navigation. Based on these findings, we provide guidelines to inform the design of future 360\degree workspace navigation techniques for large curved displays.

\end{abstract}






\maketitle

\section{Introduction}

360\degree content is becoming increasingly important and ubiquitous due to the growing demand for immersive experiences in various fields, including gaming, training simulations, and remote collaboration \cite{lee2024viewer2explorer}. 
This type of content provides users with a comprehensive and engaging way to interact with digital content, offering a sense of presence and realism that traditional media cannot match \cite{lee2024viewer2explorer}. Large curved displays offer a uniquely immersive viewing experience, particularly when displaying content such as panoramic images, videos, and `wrap-around' workspaces \cite{hsu2017holotube, lee2021viewport}. Unlike traditional flat displays, which can distort the immersive effect \cite{zannoli2017perceptual}, wall-sized curved displays offer enhanced depth perception \cite{hsu2017holotube} and better preserve image structure when viewing omnidirectional content \cite{lee2021viewport}. 
However, despite the large dimensions of wall-sized curved displays, users are often limited to viewing only a portion -- commonly up to 180\degree -- of the entire 360\degree workspace at a time. This limited viewport to the workspace leaves much of the content off-screen. 
This requires workspace navigation to bring desired off-screen content into the viewport.
Accessing off-screen content requires an effective method to navigate the workspace, which can be difficult to implement on wall-sized curved displays \cite{zannoli2017perceptual, shupp2009shaping}. When users rely on conventional navigation techniques such as mouse panning, touch dragging, or joystick steering, the sheer size of the canvas forces repeated clutching, large arm movements that break flow and induce fatigue \cite{czerwinski2003toward, hincapie2014consumed}. This mismatch between display scale and input scale highlights a pressing need for more intuitive methods that allow users to reveal unseen regions of the workspace as spontaneously as they shift their head orientation. 

Previous research on workspace navigation in wall-sized flat displays has primarily explored the use of handheld controllers \cite{tan2004physically, radle2014bigger} or mid-air hand gestures \cite{takashima2015exploring}. 
With handheld controllers, users commonly drag and flick while pressing a button on the controller to move the workspace \cite{radle2014bigger}, whereas joystick-based navigation involves pushing a joystick in a direction \cite{santos2018exploring}. 
Prior work showed that using a handheld controller or joystick-based rotation can reduce immersion \cite{stebbins2019redirecting}, cause physical strain \cite{czerwinski2003toward}, or lead to precision issues \cite{fehlberg2010active} and fatigue \cite{hincapie2014consumed}. Head movements, which users naturally use to move focus to areas of interest, have been successfully employed in CAVEs \cite{ragan2012studying, ragan2016amplified}, VR \cite{stebbins2019redirecting, noronha2012sharing}, and smart devices \cite{jannat2022face, vo2022faceui, saidi2021holobar} to enhance task performance and access off-screen content. However, their application to wall-sized (curved) displays remains unexplored. 

\update{Prior research demonstrates that the choice of mapping function, such as rate-control or flick-based scrolling, significantly impacts user performance in workspace movement and pointing tasks~\cite{zhai1996human,tsandilas2013modeless,nancel2013high,aliakseyeu2008multi,malacria2015push}.
In rate-control mappings, input magnitude directly controls workspace or cursor velocity \cite{zhai1996human,casiez2008effect}; for instance, \textit{Linear} functions ~\cite{ragan2016amplified} applied to workspace movement in CAVE environments for searching tasks and \textit{Sigmoid} functions ~\cite{nancel2013high} for cursor control on wall-sized flat displays during pointing tasks, and \textit{Polynomial} functions ~\cite{tsandilas2013modeless} for cursor movement on desktop screens.
Conversely, flick-based scrolling ~\cite{aliakseyeu2008multi} explored on smartphones using mechanics such as friction-like deceleration to gradually slow the motion, abrupt stopping to immediately halt movement, and additive speed updates that increase scrolling velocity based on an additional scrolling gesture.
Together, these studies demonstrate that both rate-control and flick/zone-control mappings can be highly effective for scrolling on smartphones~\cite{aliakseyeu2008multi}, cursor movement~\cite{tsandilas2013modeless, nancel2013high}, and scrolling or workspace movement~\cite{fashimpaur2023investigating, aliakseyeu2008multi} across a range of devices.
However, it remains unclear which of these mapping families best supports head-based workspace movement on wall-sized curved displays. Consequently, we systematically compare all seven variants to identify the most promising candidate for subsequent evaluation in realistic tasks.}

In this paper, we investigate how we can translate head movements into workspace movements to access off-screen content on large curved displays.
We tested seven such mapping functions in an abstract pointing task.
In our first study, we tested three rate control functions: linear, sigmoid, and polynomial, and four zone control functions: continuous, friction, interrupted, and additive. 
Results showed that linear, sigmoid and polynomial functions outperformed continuous, friction, interrupted, and additive across several metrics, including trial time, perceived workload (NASA TLX), VR sickness (VRSQ scores), and user preference. 
The polynomial function emerged as the best choice among the tested. It achieved the fastest trial times, received the highest subjective ratings, and was the most preferred by participants. 
\update{Having identified polynomial rate control as the most effective mapping in our first study, our second study then aimed to validate its practical effectiveness against established methods.}
We therefore compared the head-based polynomial technique against two industry-standard techniques: joystick-based \push and controller-based \drag in a realistic map navigation task. 
Results showed several advantages of the head-based technique over controller-based techniques, including faster performance, reduced effort, strong preference, and lower VR sickness, all while maintaining high accuracy. 
Participants also expressed a strong preference for the head-based technique over the controller-based methods.

We make the following contributions: 

\begin{itemize}
    \item A novel approach leveraging head movements for workspace navigation on wall-sized curved displays to access off-screen information.
    
    \item \updateNEW{Results from} an evaluation of three rate control and four zone control head-to-workspace mapping functions, revealing that among the investigated mapping functions, polynomial, sigmoid, and linear outperformed continuous, friction, interrupted, and additive in terms of trial time and user preference. 
    
    \item \updateNEW{Results from} a comparative study demonstrating that the most effective head-based mapping (polynomial rate-control) outperforms traditional controller-based techniques, highlighting its advantages over industry-standard methods.

    \item A set of design guidelines derived from two user studies, such as prioritizing polynomial rate control, avoiding clutch-heavy drag interactions, and providing joystick-based navigation as a secondary option.    
  
\end{itemize}

    

\section{Background}
We review prior studies on mapping functions for workspace navigation, interactions on wall-sized curved displays, interactions for 360\degree workspaces, and head-based techniques to improve user interaction for 360\degree workspaces.

\subsection{Mapping Functions for Workspace Navigation}
\update{Mapping functions for workspace or cursor movement are generally categorized into two styles: rate control and position control \cite{hinckley2002quantitative}. In position control, users often perform drag-and-flick actions to translate input displacement into workspace displacement \cite{fashimpaur2023investigating, aliakseyeu2008multi}.
However, for head-based interaction, continuous drag-style control can lead to neck fatigue \cite{williams2008evaluation, fouche2017head}, whereas short head flicks minimize prolonged rotation by keeping movements brief and intermittent.
Therefore, for head-based workspace navigation, we adopt rate control and flick-based mapping over position control.}

\update{\textbf{Rate control}: Here, the movement velocity is determined by the direction and magnitude of the user's input, for example, tilting a joystick to move the workspace faster - enabling continuous, low-effort traversal across large distances \cite{teather2014position, fashimpaur2023investigating, zhai1996human}. This technique is particularly effective with self-centering inputs such as controller joysticks \cite{casiez2008effect, zhai1996human}. Prior work reported mixed results depending on task demands and mapping design.
Hinckley et al. \cite{hinckley2002quantitative} found that linear rate control excelled for long-distance workspace movement on desktops, whereas drag-\&-flick interaction was better for short, precise adjustments.
Tsandilas et al. \cite{tsandilas2013modeless} evaluated several mapping functions for extending the range of wrist input, comparing rate control with a polynomial transfer function, position control, and a hybrid of both. They found that rate control is most effective when input precision is low, and the target is farther away.
On the other hand, Nancel et al. \cite{nancel2013high} showed that sigmoid mappings support both fine adjustments and rapid movement when transferring small touch inputs to large-display cursor motion. 
Rate-control principles have also been applied to scrolling. Aceituno et al. \cite{aceituno2017design} explored edge-scrolling, a technique where the workspace automatically scrolls when the user points near the window/viewport edge. They compared four mapping functions: a position-based method, polynomial rate control, slow linear rate control, and fast linear rate control. Results showed that fast linear and polynomial rate mappings produced the highest speeds but also introduced control challenges.
Antoine et al. \cite{antoine2017forceedge} introduced ForceEdge, an autoscroll technique for smartphones with force sensing, where scrolling rate is controlled by applied pressure. Results showed that ForceEdge enables faster, more controllable scrolling and reduces overshoot compared to conventional distance-based edge scrolling.}

\update{\textbf{Flick-based mapping}: Flick-based mappings involve flicking for quick, large-distance navigation \cite{aliakseyeu2008multi, quinn2013touch} and dragging for precise movements -- similar to the flick and drag interactions commonly used on smartphone touchscreens. 
Fashimpaur et al. \cite{fashimpaur2023investigating} explored leveraging wrist movements for scrolling in extended reality (XR). 
They introduced `wrist Joystick', a rate control technique, and `wrist Drag', a position control technique based on drag-\&-flick. Results showed that both techniques performed comparably, maintained effectiveness in a relaxed arm posture, and provided ergonomic benefits for scrolling tasks.
Aliakseyeu et al. \cite{aliakseyeu2008multi} studied four flick-based scrolling techniques -- addition-based, friction-based, continuous, and interrupted -- to call-in items currently not visible in the viewport. They studied usage on a PDA, a TabletPC, and a digital tabletop. Across devices, interrupted flicking was the most effective for scrolling. Malacria et al. \cite{malacria2015push} introduced Push-edge and Slide-edge scrolling, designed to overcome limitations of traditional rate-based edge scrolling by enabling users to actively ``push'' against the viewport edge. Push-edge relies on position control, while Slide-edge incorporates simulated inertia to support longer movements. Both techniques outperformed standard edge scrolling -- reducing overshoot and yielding up to 13\% faster selection times (without increasing errors).}

\update{Rate-based and flick-based functions have been studied for desktops, smartphones, and XR environments; their effectiveness for head-based interaction on large displays is unknown. This motivates our empirical work, where we compare the two function types when used for head-driven workspace navigation.}

\begin{figure*}[!ht]
    \centering
    \includegraphics[width=0.90\textwidth]{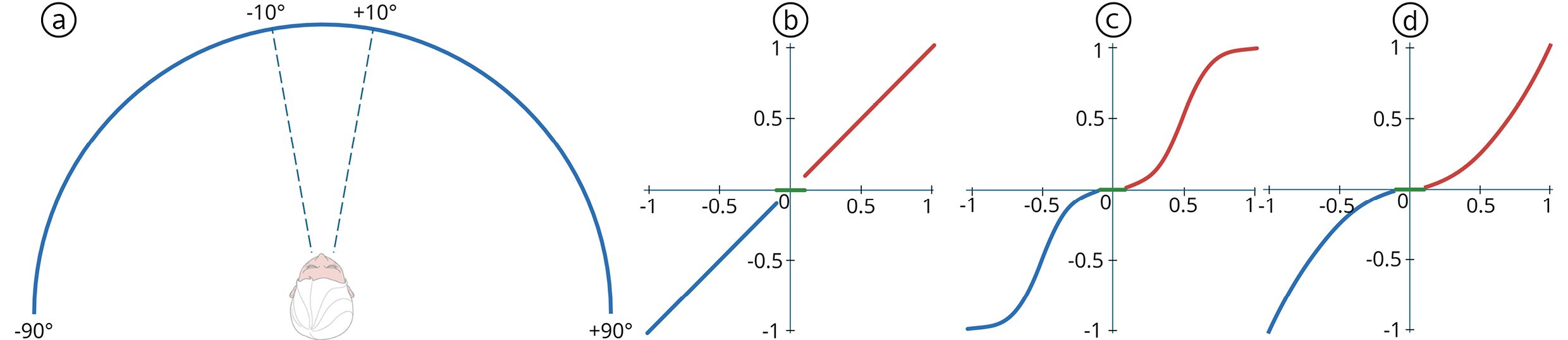}
	\caption{(a) A user can control the velocity of the workspace movement by rotating the head in the horizontal direction. The workspace velocity is set to zero between $\pm10$\textdegree, allowing the user to stop moving the workspace. Rate control mapping function (b) Linear, (c) Sigmoid, (d) Polynomial where the vertical axis represents the velocity and the horizontal axis the head rotation.   }
    \Description{Figure 1 contains four parts: (a), (b), (c), and (d). The figure illustrates different rate control methods for manipulating the workspace using head movements on a large curved display. (a) shows a user controlling the velocity of the workspace movement by rotating their head horizontally. The horizontal axis, or yaw, represents the user's head rotation from -90° (far left) to +90° (far right). A specific stop zone is marked between -10° and +10°, where the velocity of the workspace is set to zero. This allows users to stop moving the workspace entirely when their head rotation falls within this range. The velocity increases as the user's head moves outside of this range, allowing for faster movement of the workspace. (b) shows a linear rate control mapping function. The horizontal axis represents head rotation (x), normalized from -1 (looking far left) to +1 (looking far right), and the vertical axis represents workspace velocity (y), normalized from -1 (maximum left velocity) to +1 (maximum right velocity). In this mapping, the velocity increases uniformly as the head rotation increases or decreases. The workspace movement follows a linear relationship between head rotation and velocity. (c) shows a sigmoid rate control mapping function. The vertical axis represents workspace velocity, and the horizontal axis represents head rotation. In this mapping, workspace velocity follows a sigmoid curve, where small head rotations correspond to lower velocities, and larger head rotations correspond to higher velocities. The workspace velocity accelerates quickly as the head moves further from the center, reaching a moderate speed range around 0.5. (d) shows a polynomial rate control mapping function. The vertical axis represents workspace velocity, and the horizontal axis represents head rotation. The polynomial mapping results in slower workspace movement for smaller head rotations, with a rapid increase in speed for larger head rotations. This is useful for precise control when users make small head movements and faster navigation when head movements are larger. Each rate control mapping function translates head movement into a corresponding velocity that moves the workspace left or right, depending on the direction of the user's head rotation. The stop zone allows for halting the workspace movement entirely when the user's head is within the designated range.}
	\label{fig:rate_control} 
\end{figure*}

\subsection{Interaction with 360\degree Workspaces}

Navigating 360\degree video and immersive environments presents a fundamental challenge: the user's field of view (FOV) is inherently limited, causing important content to often fall outside the viewport \cite{lee2024viewer2explorer, noronha2012sharing}. This limitation forces users to constantly scan the environment, risking that they might miss key events while looking in the wrong direction \cite{pavel2017shot}. The problem is magnified in scenes with multiple regions of interest, where users must divide their attention and may become disoriented \cite{lin2017outside}. Moreover, continuously rotating the head to reveal off-screen content can quickly lead to fatigue \cite{stebbins2019redirecting}; this issue highlights the critical importance of having a suitable mapping function, as prior work has often relied on basic linear mappings. 

To address these issues, researchers have explored both automated and user-guided solutions. One line of research focuses on automatically orienting the workspace to ensure important content remains visible. For example, Pavel et al. \cite{pavel2017shot} proposed workspace-oriented cinematography techniques that reorient 360\degree video at each cut to align important content with the user’s FOV. Similarly, saliency-aware systems compute smooth camera paths that follow prominent objects, creating traditional ``flat'' video experiences from spherical content \cite{kang2019interactive}. While such methods reduce manual navigation, they risk limiting user agency or causing discomfort if camera movements are overly pronounced \cite{kang2019interactive, kolasinski1995simulator}. Stebbins and Ragan \cite{stebbins2019redirecting} evaluated an automated workspace adjustment in a 360\degree video player and found that slower scene rotations reduced physical effort without being disruptive. 

Researchers also provided contextual cues for off-screen content while preserving user control. Arrow indicators are a simple solution, but they often confuse users by showing direction without revealing content. To improve awareness, Lin et al. \cite{lin2017outside} introduced Outside-In, which displays off-screen regions of interest as picture previews at the viewport edge. Compared to arrows, this helped users to better understand spatial relationships and storylines. Kang and Cho \cite{kang2019interactive} presented a hybrid system that combines saliency-aware automatic navigation with real-time user adjustments, improving comfort over purely manual navigation. Lee et al. \cite{lee2024viewer2explorer} also demonstrated Viewer2Explorer, a map-based desktop interface for 360\degree museum videos, which significantly enhanced spatial awareness, autonomy, and engagement compared to timeline-based browsing. 

While prior work has focused on automatic camera control and visual cues, less attention has been paid to the underlying mapping functions. Specifically, there is a lack of research into how to best leverage natural head movements for navigating 360\degree content.

\subsection{Workspace Navigation using Head and Gaze Input}
\update{Head movements have been widely investigated as an input modality for desktop and immersive environments. On desktop platforms, continuous head-based interactions have been explored for fine-grained control. 
For instance, Nukarinen et al.'s HeadTurn~\cite{nukarinen2016evaluation} allows users to adjust numeric values by turning their head left or right, while Tang et al.'s HeadPager~\cite{tang2016headpager} enables faster page navigation compared to mouse scrolling using head-turns beyond a threshold. 
Similarly, Jacob et al.~\cite{rotationaldesktop} demonstrated that orbital camera control using head rotations provides sufficient precision (1\textdegree) for fine-grained tasks, and Teather and Stuerzlinger~\cite{teather2008exaggerated} showed that amplifying head-coupled motions can enhance camera navigation efficiency in desktop VR games, even outside traditional HMD or CAVE setups.}

\update{In immersive environments, such as HMD and CAVE setups, head movement along the horizontal (yaw) is the primary way users explore 360\degree workspaces in VR~\cite{stebbins2019redirecting, noronha2012sharing,ragan2012studying, ragan2016amplified}. 
However, extensive head rotation can lead to physical fatigue and reduce engagement~\cite{lee2024viewer2explorer, noronha2012sharing, stebbins2019redirecting, ragan2016amplified}. 
Consequently, researchers investigated methods for leveraging and scaling head movements to mitigate these challenges. For instance, Razzaque et al.~\cite{razzaque2002redirected} proposed redirected walking, which gradually adjusts the virtual scene orientation based on head and torso rotation, allowing traversal of large virtual spaces within a limited physical area. Langbehn et al.~\cite{langbehn2019turn} evaluated dynamic rotation gains (head rotation scaled by parabolic gains), static gains (scaled by a constant), scrolling (constant gain), and swivel-chair rotation for users with limited physical turning capabilities. Dynamic gains outperformed the other approaches in terms of VR sickness, usability, and preference. Westhoven et al.~\cite{westhoven2016head} examined dampening (0.90), control (1.0), and boosting (1.10) rotation factors in a target-acquisition task, showing that dampening and boosting increased perceived workload. Tregillus et al.~\cite{tregillus2017handsfree} studied using head-tilt for hands-free navigation in mobile VR. They compared pure head-tilt (TILT), head-tilt combined with walking-in-place movements (WIP-TILT), and joystick use. TILT enabled faster navigation with fewer collisions, and both TILT and WIP-TILT were rated higher in presence while causing similar levels of cybersickness as joystick control.}

\update{In CAVE environments, mapping head rotation to workspace navigation has been studied. Freitag et al.~\cite{freitag2016examining} found that higher rotation gains reduced spatial knowledge without increasing simulator sickness. Similarly, Ragan et al.~\cite{ragan2012studying} showed that head-based navigation improved spatial judgment accuracy compared to controller-based navigation. In a follow-up study~\cite{ragan2016amplified}, they demonstrated that varying rotation gains had little effect on accuracy or task time, and that simulator sickness was higher with HMDs than in CAVEs. Furthermore, McGill~\cite{mcgill2020expanding} explored mapping head rotations to virtual display positions and showed that ``dead zones'' on the displays effectively reduced neck fatigue and discomfort.}

\update{Gaze-interaction has also been explored. 
Lee et al.~\cite{lee2024snap} showed that gaze-input enables efficient 360\degree navigation in VR, with performance comparable to head-based and controller-based methods.
Pai et al.~\cite{pai2017gazesphere} combined gaze with head movements, letting users select a navigation direction by looking at virtual cues and then turning their head to move smoothly along pre-recorded 360\degree transition videos. Study participants were able to quickly learn the technique and reported experiencing minimal or no motion sickness.}

\update{Overall, previous work highlights the versatility of head-based interaction -- alone or in combination with gaze -- for precise, efficient, and hands-free workspace navigation in desktop, VR, and CAVE environments. However, little is known about how head-based interaction techniques can be used for workspace navigation on wall-sized curved displays, where physical constraints and display geometry may influence navigation strategies and user performance. 
Accordingly, we studied head-based interaction for such settings, focusing on methods to map head movements to workspace movements and evaluating their impact on user efficiency.}
\section{Central Factors for accessing off-screen information}

We explored three central factors for accessing off-screen information via head movements on a large curved display: Mapping functions, display window size, and target distance.

\begin{figure*}[!h]
    \centering
    \includegraphics[width=0.93\textwidth]{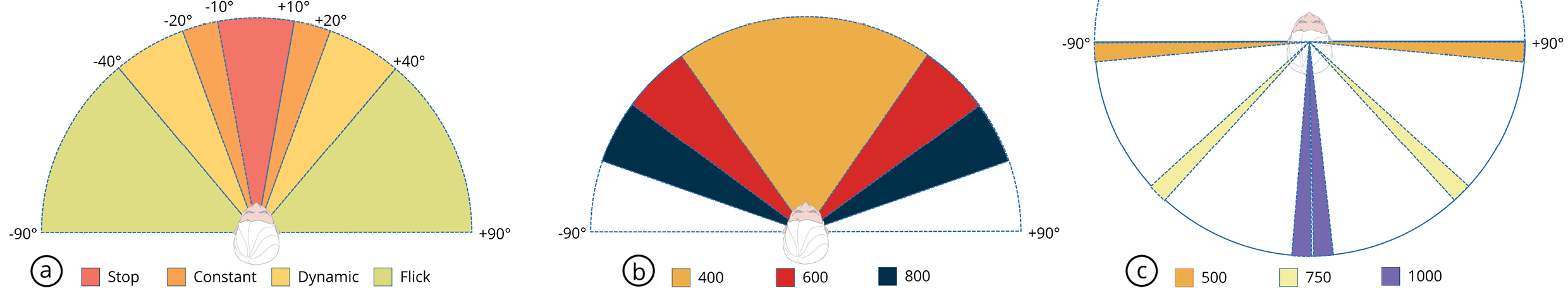}
	\caption{(a) Four zones: Stop, Constant, Dynamic, and Flick, for the zone control techniques; (b) three different display window sizes (400cm, 600cm, and 800cm); and (c) three different target distances (500cm, 750cm, and 1000cm) used in the user study.} 
    \Description{Figure 2 has three sections (a to c). In (a), the semi-circular diagram illustrates the four different zones for zone control techniques based on head rotation. The zones are color-coded: the red center region represents the Stop zone, where workspace velocity is set to zero between -10° and +10°; the orange sections represent the Constant zone, where workspace velocity is slow and steady between -20° and -10° (left) and +10° and +20° (right); the yellow sections represent the Dynamic zone, where workspace velocity is higher, ranging from -40° to -20° and +20° to +40°; and the light green sections on the edges represent the Flick zone, where workspace movement is initiated when the user's head rotates beyond -40° and +40°. In (b), the semi-circular diagram depicts three different display window sizes, showing how much content can be viewed at once: the yellow section corresponds to a 400cm display window, the red section corresponds to a 600cm display window, and the dark blue section corresponds to an 800cm display window. Finally, in (c), the semi-circular diagram displays three different target distances, indicating the proximity of targets to the user: red indicates a 500cm distance, yellow indicates a 750cm distance, and blue indicates a 1000cm distance.}
	\label{fig:position-and-experiment} 
\end{figure*}

\subsection{Mapping Functions}  
For mapping functions based on head movement, we focused solely on the users' horizontal head rotation angle (i.e., yaw) relative to the center of the curved display. With yaw, users need to turn their heads left or right, which is more intuitive and less tiring than incorporating pitch, i.e., up and down movements, or roll, i.e., tilting the head sideways \cite{ragan2016amplified}. We thus investigated the following \textit{Mapping Functions}. 

\subsubsection{Rate control} With rate control, the workspace rotates at a velocity determined by the user's head rotation angle. For instance, when users rotate their heads at a smaller angle from the center of the display, the workspace moves slowly. As the angle increases, the workspace moves faster. As suggested by Fashimpaur et al. \cite{fashimpaur2023investigating}, we reserved a specific region for all our rate control mapping functions ($\pm 10\degree$) to allow the users to stop the workspace movement. We called this area a `stop' zone where the velocity is set to zero (Figure \ref{fig:rate_control}a). \update{This zone is critical for mitigating the ``Midas Touch'' problem, where unintended head movements (e.g., casual glancing) could otherwise trigger unwanted workspace rotation. By requiring a deliberate head turn beyond $\pm 10\degree$, we prevent accidental activations.}

We used a curved display (see Apparatus for hardware details) with a total viewing angle of 180\textdegree. Therefore, we set the maximum horizontal head rotation to the right to be +90\textdegree$~$ and the maximum horizontal head rotation to the left to be -90\textdegree. 
As horizontal head movement is the input ($x$), we normalize the input range by mapping [-90\textdegree, +90\textdegree]$~$ to $[-1, 1]$. Here, $x=-1$ corresponds to a -90\textdegree$~$ head rotation to the left, $x=1$ corresponds to a +90\textdegree$~$ rotation to the right, and $x=0$ is the center. The output of our mapping functions is the corresponding workspace velocity ($y$). The output velocity is also normalized to $[-1, 1]$, where $y=1$ corresponds to the maximum rightward scrolling speed of 100\textdegree/sec (i.e., 570.56 cm/sec) and $y=-1$ corresponds to the maximum leftward speed of -100\textdegree/sec (i.e., -570.56 cm/sec).

\textbf{Linear}: The linear function creates a consistent mapping where the velocity increases uniformly with the degree of horizontal head rotation based on Equation \ref{eqn:linear}, resulting in a direct, proportional relationship between head rotation and workspace velocity (Figure \ref{fig:rate_control}b). 
\begin{equation}
y =
\begin{cases}
x & \text{if } |x| > 0.11 \; (\text{i.e., }x <-10\degree \text{or } x > 10\degree\rule{-0.25em}{0.0em})\\  
0 & \text{if } |x| \leq 0.11 \; (\text{i.e., }-10\degree\rule{-0.25em}{0.0em}\le x \le 10\degree\rule{-0.25em}{0.0em})
\end{cases}
\label{eqn:linear}
\end{equation}



\textbf{Sigmoid}: The sigmoid function maps lower head rotation values to lower workspace velocities and higher rotation values to higher workspace velocities. Here, most workspace velocity values are mapped around the middle range (near 0.5 in Figure \ref{fig:rate_control}c), resulting in moderate speeds for head rotations that are neither extreme nor minimal. With a pilot study, we settle on $p=10$, which determines the shape of the sigmoid function and \textit{offset = 5}, which shifts the center of the function, as the optimal values for our use cases in Equation \ref{eqn:sigmoid}:
\begin{equation}
y =
\begin{cases}
\frac{1}{1 + e^{-x \cdot p + offset}} & \text{if } x > 0.11  \;(\text{i.e., }x > 10\degree\rule{-0.25em}{0.0em})\\
-\frac{1}{1 + e^{x \cdot p + offset}} & \text{if } x < -0.11 \;(\text{i.e., }x < -10\degree\rule{-0.25em}{0.0em})\\
0 & \text{if } |x| \leq 0.11 \; (\text{i.e., }-10\degree\rule{-0.30em}{0.0em}\le x \le 10\degree\rule{-0.25em}{0.0em})
\end{cases}
\label{eqn:sigmoid}
\end{equation}

\textbf{Polynomial}: 
The polynomial function results in very slow workspace movement for small head rotations and a rapid increase in speed for larger head rotations from the center of the screen (Figure \ref{fig:rate_control}d). For the polynomial function, we used Equation \ref{eqn:polynomial} where b is the exponent that determines the shape and behavior of the graph. When $b=1$, the function is linear, but when $b>1$, the graph becomes nonlinear. 
Through a pilot study, we found that $b=2$ is the optimal choice for our use case, as also used in \cite{tsandilas2013modeless}. 
\begin{equation}
y =
\begin{cases}
\text{sign}(x) \cdot |x|^b & \text{if } |x| > 0.11 \; (\text{i.e., }x <-10\degree \text{or } x > 10\degree\rule{-0.25em}{0.0em})\\
0 & \text{if } |x| \leq 0.11 \; (\text{i.e., }-10\degree\rule{-0.25em}{0.0em}\le x \le 10\degree\rule{-0.25em}{0.0em})
\end{cases}
\label{eqn:polynomial}
\end{equation}

\subsubsection{Zone control} 

While traditional position control maps user movement directly to displacement, applying this paradigm to head-based interaction would require constant, repetitive head motion to move the workspace, which could cause neck strain and fatigue. 
We investigate the performance of \textit{zone control}, which divides the interaction space (i.e., the display window) into discrete zones where entering a zone triggers an action, with the option to perform flick or use rate control inside zones. 
Instead of continuously tracking the head movement, zone control enables short, flick-like gestures -- similar to touchscreen flicks -- for efficient navigation.

We investigated four zone control methods -- continuous, friction, additive, and interrupted mapping.  
Here, the 180\textdegree~display, or display window, was divided into stop, constant, dynamic, and flick zones (see Figure \ref{fig:position-and-experiment}a), which we discuss below. 
Similar to the rate control mapping functions, the workspace velocity is normalized in [-1, 1] using the following formula:
\begin{equation}
y =
\begin{cases}
\text{sign}(x) \cdot \frac{\max(0, maxTime - t)}{maxTime} & \text{if } |x| > 0.44 \; \text{(Flick:} |x| > 40\degree\rule{-0.25em}{0.0em)} \\
\text{sign}(x) \cdot c & \text{if } 0.11 < |x| \leq 0.22 \; \\& \text{(Constant:} 20 \degree\rule{-0.35em}{0.0em}< |x| \le 40 \degree\rule{-0.25em}{0.0em}) \\
0 & \text{if } |x| \leq 0.11 \; {\text{(Stop:} |x| \leq 10\degree\rule{-0.25em}{0.0em})}
\end{cases}
\label{eqn:positionALL}
\end{equation}
where $c$ is the constant speed in the Constant Zone, and $t$ is the time in seconds spent in the Dynamic Zone.

\begin{itemize}
  \item \textbf{Stop Zone}: Like the rate control techniques, we have a `Stop' zone where the workspace velocity is zero. Users are within the stop zone when the horizontal head rotation angle is between $\pm10$\degree (i.e., $-0.09<x<0.09$ ). 
  
  \item \textbf{Constant Zone}: We observed that the stop zone was insufficient for the mappings, it was difficult to stop the movement of the workspace abruptly without first slowing it down. To address this, we introduced the `Constant' zone with constant velocity of $y=\pm 0.10$ (10\textdegree/sec or 57.07 cm/sec), where the workspace moves with a slow and steady speed when the user's head is within $0.11< |x| < 0.22$ (i.e., $\pm$10\degree to $\pm$20\degree for slow movement towards the right and left) (Figure \ref{fig:position-and-experiment}a).

  \item \textbf{Dynamic Zone}: To explore the impact of different mapping functions (e.g., continuous, friction, additive, or interrupted, as discussed below), we defined a `Dynamic' zone where each function exhibits its unique behavior. For instance, in a certain mapping (e.g., friction mapping), the workspace velocity decreases gradually within the Dynamic Zone. To define this zone, we used a head angle of +20\degree to +40\degree and -20\degree to -40\degree (i.e., $0.22< |x| < 0.44$), on the right and left side of the screen, respectively. 

  \item \textbf{Flick Zone}: 
  The `Flick' zone is designated to initiate rapid workspace movement with head rotation for the four mapping functions.
  To start a flick, the user moves the head from the dynamic zone to the flick zone in Figure \ref{fig:position-and-experiment}a (i.e., $|x| > 40\degree$\rule{-0.25em}{0.0em}). The workspace velocity for a flick is determined by how long the user spends in the dynamic zone before entering the flick zone; we denote this time as `t'. In Equation \ref{eqn:positionALL}, we see the maximum time (`t' = ``maxTime'') leads to a speed of 0, whereas the lower the values of $t$, the faster the flick speed, where the maximum flick speed is 1 (100\textdegree/sec or 570.56 cm/sec) for $t=0$. Through pilots, we set ``maxTime'' for a flick to 2 seconds. 
\end{itemize}

All zone parameters were optimized through pilot studies to ensure the best range of head rotation values. A circle cursor, based on the user's head direction, helps visualize head pointing. In all four mappings, users can stop workspace movement by moving the cursor to the stop zone or keep a constant speed in the constant zone. In the dynamic zone, workspace velocity changes based on the mapping function, and the time spent there before entering the flick zone determines the flick speed.

\begin{figure*}[!ht]
    \centering
    \includegraphics[width=1.0\textwidth]{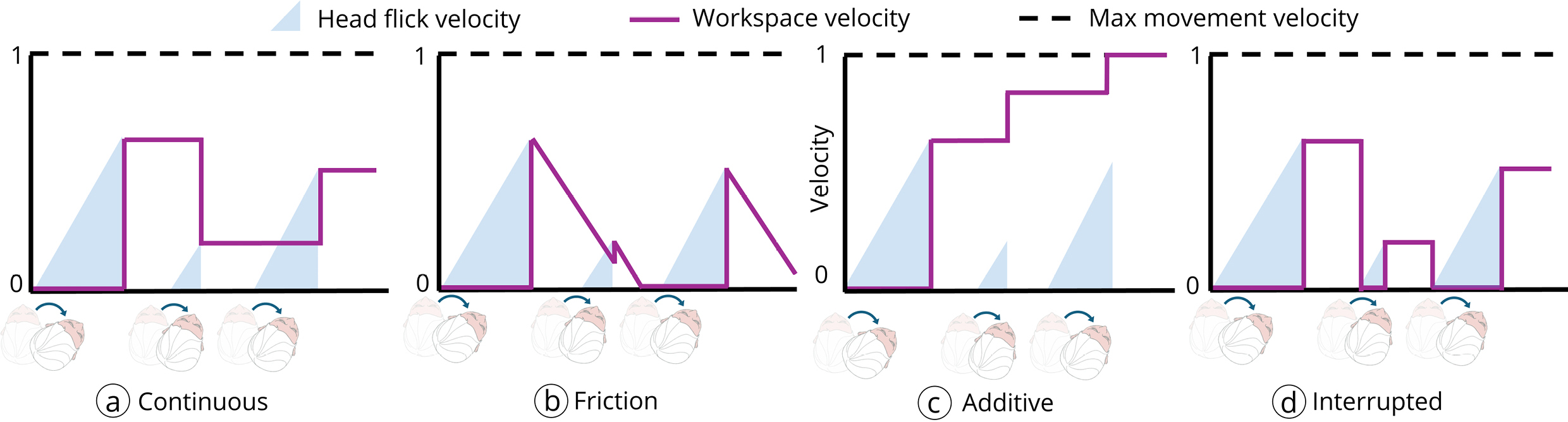}
	\caption{Zone control techniques mapping Function (a) Continuous, (b) Friction, (c) Additive, (d) Interrupted where the vertical axis is the velocity of the workspace and the horizontal axis shows head movement performed by the user.}
    \Description{The figure 3 showcases four different flick-based mapping functions, each depicted as a graph, with the vertical axis representing workspace velocity (ranging from 0 to 1) and the horizontal axis representing the user's head movement. Illustrations of head rotations are shown below each graph, depicting how head movements affect workspace velocity. In (a) Continuous, the workspace velocity is updated and maintained based on the user's head flick, staying constant while the head remains in a specific zone. In (b) Friction, the workspace velocity decreases gradually, simulating friction that slows down the movement. In (c) Additive, multiple consecutive head flicks increase the workspace velocity by adding to the current speed. In (d) Interrupted, the workspace movement stops abruptly when the user's head enters a designated region, reducing velocity to zero.}
	\label{fig:pos-mapping} 
\end{figure*}

We explored four mapping functions: Continuous, Friction, Additive, and Interrupted. They share the same fundamental zone structure and flick-initiation mechanism. Their primary difference lies in how they manage the workspace movement velocity when the user's head is in the Dynamic Zone after a flick, as described below.

\textbf{Continuous}: In the continuous mapping, when the user performs a flick, the workspace velocity is updated based on the flick. The user can maintain this speed as long as the cursor remains within the constant or dynamic zone, see Equation \ref{eqn:Continuous}. To adjust the speed again, the user can perform a new flick, which sets a new velocity for the workspace, as shown in Figure \ref{fig:pos-mapping}a.
\begin{equation}
y_{\text{new}} = y_{\text{current}} \quad \text{if } 0.22 < |x| \leq 0.44 \,\,(\text{i.e., }20 \degree\rule{-0.5em}{0.0em}< |x| \le 40 \degree\rule{-0.25em}{0.0em})
\label{eqn:Continuous}
\end{equation}

\textbf{Friction}: This mapping is similar to Continuous mapping; however, it includes an additional coefficient $\mu$ (in Equation \ref{eqn:Friction} set to 0.03 with pilot testing) determines how long the workspace takes to reduce speed to the point of stopping, and `$t_2$' is the elapsed time in seconds within the Dynamic Zone (see Figure \ref{fig:pos-mapping}b). 
We incorporate this mapping as it is commonly used for workspace navigation on desktop and mobile devices, where the navigation speed decreases gradually, mimicking the effect of friction until the movement eventually comes to a stop \cite{aliakseyeu2008multi}.
\begin{equation}
y_{\text{new}} = y_{\text{current}} - \text{sign}(y_{\text{current}}) \cdot \mu \cdot t_2 \quad 
\begin{aligned}[t]
&\text{if } 0.22 < |x| \leq 0.44 \\
&(\text{i.e., } 20 \degree < |x| \le 40 \degree)
\end{aligned}
\label{eqn:Friction}
\end{equation}

\textbf{Additive}: The additive mapping function is similar to the continuous mapping function, with the key difference being that users can increase the workspace movement through additional flicks. Figure \ref{fig:pos-mapping}c shows the additive function (in Equation \ref{eqn:additive}) where the workspace velocity 1 is achieved by performing consecutive flicks.
\begin{equation}
y_{\text{new}}=
\begin{cases}
\text{sign}(x) \cdot \min(1, | y_{\text{current}} +  \frac{\max(0, maxTime - t)}{maxTime}|) \\\,\,\,\,\,\,\,\,\,\,\,\,\,\,\,\,\,\,\,\, \text{if } |x| > 0.44 \,\, (\text{i.e., }x > 40 \degree \text{or } x<-40\degree\rule{-0.25em}{0.0em})\\ 
y_{\text{current}} \\\,\,\,\,\,\,\,\,\,\,\,\,\,\,\,\,\,\,\,\,\text{if } 0.22 < |x| \leq 0.44 \,\,  (\text{i.e., }20 \degree\rule{-0.5em}{0.0em}< |x| \le 40 \degree\rule{-0.25em}{0.0em}) \\
\end{cases}
\label{eqn:additive}
\end{equation}

\textbf{Interrupted}: The interrupted mapping function (Equation \ref{eqn:Interrupted}) is also similar to the continuous mapping function, with the key difference being that whenever the cursor enters the dynamic region, the workspace movement velocity is reduced to zero, allowing the users to stop quickly \cite{aliakseyeu2008multi} (Figure \ref{fig:pos-mapping}d). 
\begin{equation}
y = 0 \quad \text{if } 0.22 < |x| \leq 0.44 \,\,(\text{i.e., }20 \degree\rule{-0.5em}{0.0em}< |x| \le 40 \degree\rule{-0.25em}{0.0em})
\label{eqn:Interrupted}
\end{equation}

\subsection{Display Window} 
We define the display window as the area on the display in which the workspace is visible. The maximum viewing angle of our display is 180\degree \update{(i.e., 1027 cm arc length)} and the 360\degree workspace is circular with a fixed radius of 3.27 meters. However, the size of the display can vary in real-world usage scenarios. Accordingly, to generalize our findings for various scenarios, we choose three different display window sizes -- 400cm, 600cm, and 800cm, as shown in Figure \ref{fig:position-and-experiment}b. \update{These sizes corresponds to horizontal visual angle of 70.09\textdegree, 105.13\textdegree, and 140.17\textdegree, respectively. 
Researchers using different display sizes or viewing distances can replicate our setup by matching the horizontal visual angles of our display windows. For curved displays, the horizontal visual angle in degrees is
$\theta = (w/r)\cdot 180/\pi$, where $w$ is the window arc length along the surface of the display and $r$ is the display radius.}
Only the portion of the content that is inside the display window is visible to the user.

\subsection{Target Distance} 
We define target distance as the distance between the center of the display and the center of the target along the curved 360\degree display.
Having different target distances will ensure that the proposed techniques are not biased for a particular distance \cite{hinckley2002quantitative, fashimpaur2023investigating, Ullah2023Pointing}. 
We selected three target distances: 500cm, 750cm, and 1000cm (Figure \ref{fig:position-and-experiment}c). \update{At our 3.27m viewing radius, these correspond to angular separations of approximately 87.61\textdegree, 131.41\textdegree, and 175.22\textdegree $~$along the 360$^{\circ}$ workspace.} 
Our largest display window measures 800cm. When centered on the display, it has 400cm of unused display space on both sides. Accordingly, choosing a minimum target distance of 500cm ensures that targets are initially always outside the display window.

\begin{figure*}[!ht]
    \centering
    \includegraphics[width=1.0\textwidth]{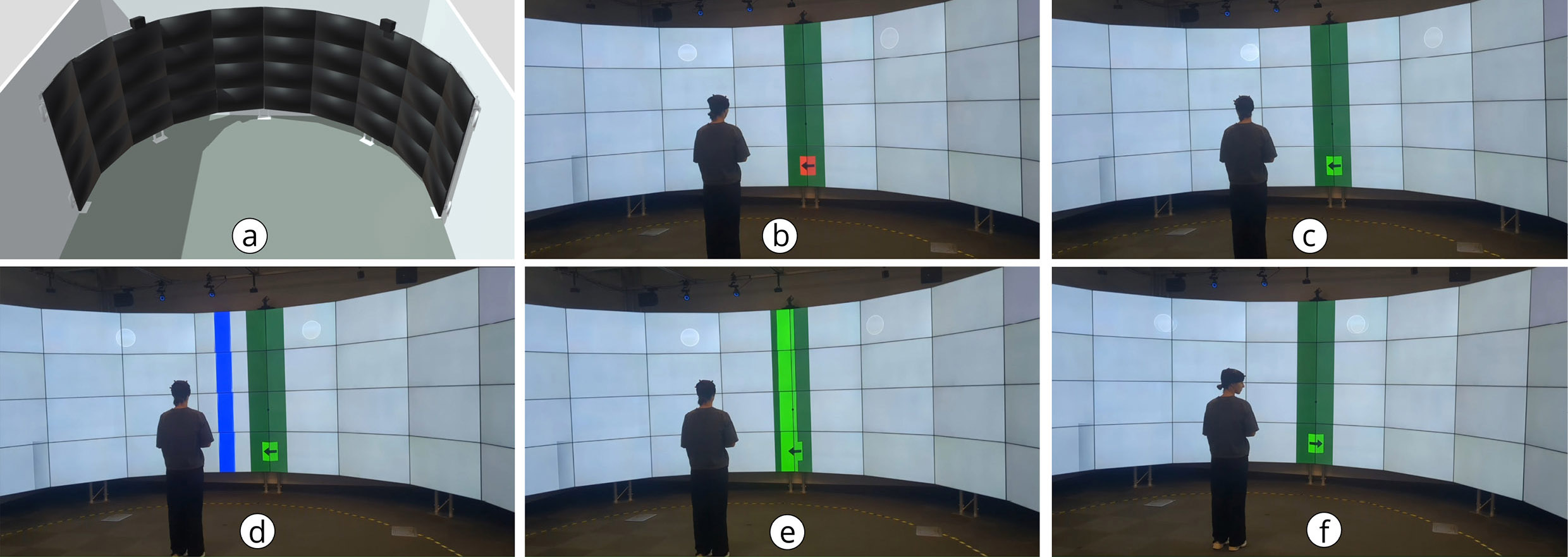}
	\caption{Study 1. (a) 3D model showing the curved display hardware setup. (b-f) Study tasks for polynomial mapping function.} 
    \Description{Figure 4 shows the study setup for a large curved display environment. (a) presents a 3D model of the large curved display hardware setup, which is used to display a 360° workspace. (b) to (f) depict the various steps involved in the study tasks when using the polynomial mapping function. In (b), the participant is positioned in front of the display, interacting with an off-screen target. The participant's head movements are tracked to bring the target on-screen, guided by visual cues. In (c) and (d), the target is aligned within the center frame. (e) shows the correct positioning of the target, and (f) marks the completion of the task.}
	\label{fig:study_1_task} 
\end{figure*} 
\section{Study 1 -- Mapping functions}
Our first study examines the mapping functions for head movement to bring off-screen information onto the display window, using different display window sizes and target distances. Our main research questions are:

\begin{itemize}
    \item \textbf{RQ1}) What is the most effective mapping function for bringing off-screen information into the display window when using head movement on large curved displays?
    
    \item \textbf{RQ2}) How does the display window size affect task performance when accessing off-screen on large curved displays?

    \item \textbf{RQ3}) How does the target distance affect the workspace navigation efficiency when interacting with off-screen information on large curved displays with head movement?
    
\end{itemize}

\subsection{Apparatus}
We conducted our study on a semi-circular large curved display ($180^\circ$ viewing angle, 3m tall and 3.27m radius) from Mechdyne~\cite{mechdyne-new1}, consisting of forty 46-inch LED 3D displays in a 4$\times$10 grid (see Figure \ref{fig:study_1_task}a). 
We used an 8-camera OptiTrack motion capture system \cite{optitrack} to track users' head movement direction with a Mocap Beanie \cite{Beanie}. Participants used the trigger button on the Meta Quest 2 controller to initiate workspace movement and confirm task completion.
The study application was created using Unity 3D with C\#. 

\subsection{Participants}
We recruited 28 participants (20 males and 8 females) from the local university, with an average age of 23.14 years (SD = 1.73). Thirteen participants had used large wall-sized displays at least once in the last 12 months, while seven had never used them. Each participant received a \$15 honorarium for their participation.
\begin{figure*}[!bht]
    \centering
    \includegraphics[width=1.0\textwidth]{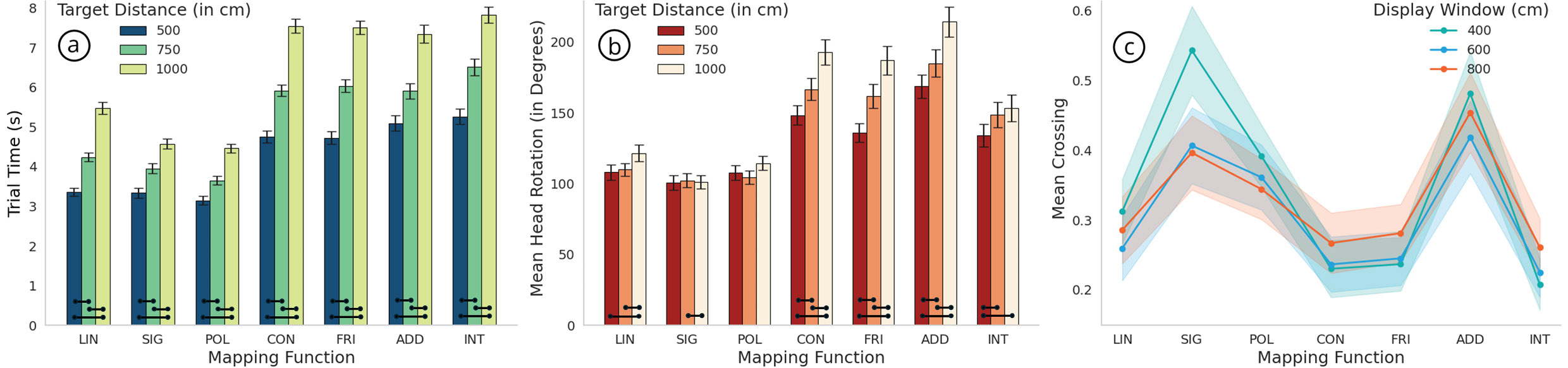}
	\caption{(a) Trial time by \textit{Mapping Function} for each \textit{Target Distance}, (b) Total head movement time by \textit{Mapping Function} for each \textit{Target Distance} and (c) Crossing by \textit{Mapping Function} for each \textit{Display Window}. Error bars represent  95\% confidence intervals. \update{Capped horizontal bars denote significance.}} 
    \Description{Figure 5 has three graphs for mapping functions, including Linear (LIN), Sigmoid (SIG), Polynomial (POL), Continuous (CON), Friction (FRI), Additive (ADD), and Interrupted (INT), for different target distances (500 cm, 750 cm, 1000 cm) and display windows (400 cm, 600 cm, 800 cm). In the first graph (a), the trial time is displayed for each mapping function, and we see that for all distances, the Additive and Interrupted mappings take significantly longer compared to the others. Polynomial, Sigmoid, and Linear functions consistently show faster times, with Polynomial showing the best performance overall. In the second graph (b), total head movement, measured in degrees, shows a similar trend, with Polynomial and Sigmoid requiring less head movement than Additive and Interrupted functions. Lastly, in the third graph (c), mean crossing data for each display window are presented. The 400 cm window shows the highest crossing values, especially for the Sigmoid and Additive mappings, while other mappings like Continuous and Friction remain relatively consistent across different window sizes. Error bars in all graphs represent 95\% confidence intervals, indicating the variation within the data.}
	\label{fig:study1_a} 
\end{figure*}

\subsection{Tasks and Procedure}

The primary task of the user study is to bring off-screen targets on-screen using various mapping functions. We used a one-dimensional navigation task inspired by previous research \cite{hinckley2002quantitative, cao2008peephole, fashimpaur2023investigating, aliakseyeu2008multi}, where users rotated their head to bring an off-screen target to the center of the visible display area. The OptiTrack system was used to capture users' head's orthogonal pointing direction to determine the horizontal (yaw) head rotation angle from the center of the display. A cursor indicated the user's pointing location for zone control mappings to assist with flick visualization, while no cursor was used for rate control mappings.

Participants were invited to the room with the large curved display. They were given an overview of the study, fitted with an OptiTrack Mocap Beanie, and took a position at the center of curvature of the screen (approx. 3.27 meters away from the center of the screen).
Their task was to locate an off-screen blue target and bring it onto the screen using head movements. A 70cm wide dark green frame, referred to as the ``center frame'' was located at the center of the screen to ensure that all off-screen targets were positioned at the center when selected. 
A 20cm square-shaped indicator with an arrow positioned at the center frame directed users in the target direction (Figure~\ref{fig:study_1_task}b). 
Off-screen targets, 30cm wide, were positioned at varying distances on either side. 
To help participants perceive workspace movement, we placed five circles with a 20cm radius along the top of the 360\degree workspace at 60\degree intervals. Once a target was brought into the center frame, participants could select it by pressing the trigger button on the controller.

To start moving the workspace using head movements, participants had to hold the trigger button for more than 300 milliseconds (a ``long press''). \update{This deliberate activation long press serves as another key mitigation against the ``Midas Touch'' problem, ensuring that workspace movement only occurs when explicitly intended and not from natural head movements during observation.} The square-shaped indicator turned green to signal that a long press was being performed; otherwise, it remained red (Figure \ref{fig:study_1_task}b). Upon detecting the long press, the workspace moved according to the participant's head movements and the mapping function (Figure~\ref{fig:study_1_task}c). When the blue target appeared within the display window, the participant needed to position it within the center frame as shown in Figure~\ref{fig:study_1_task}d. Once the blue target was fully within the center frame (Figure~\ref{fig:study_1_task}e), the target turned green, signalling correct positioning. The participant could then confirm the selection with a ``short press'' (pressing and releasing the trigger in less than 300 milliseconds), which ended the trial. A successful sound was played if the target was correctly positioned within the center frame during the short press; otherwise, a failure sound was played, and the trial continued until the task was completed successfully. After successfully completing the task, the application positioned the next off-screen target and pointed the square-shaped indicator in the new target direction, turning it red again. When ready, the participant could now start the next trial (Figure \ref{fig:study_1_task}f).


\subsection{Design}

We used a within-subjects design with independent variables - \textit{Mapping Function}: Linear (Lin), Sigmoid (Sig), Polynomial (Pol), Continuous (Con), Friction (Fri), Additive (Add), Interrupted (Int), \textit{Display Window}: 400, 600, and 800cm and \textit{Target Distance}: 500, 750, and 1000cm. The order of \textit{Mapping Function} was counter-balanced across participants using a Latin square. For each \textit{Mapping Function}, a \textit{Display Window} is randomly selected. For a specific \textit{Display Window}, the participant had to perform 24 trials, consisting of 8 repetitions of 3 target distances, which appeared in random order. We ensured that each target distance appeared on the left side half of the time and on the right side half of the time. The study design was: \textit{Mapping Function} (7) $\times$ \textit{Display Window} (3) $\times$ \textit{Target Distance} (3) $\times$ repetition (8) = 504 trials per participant. 
Participants had ten practice trials before starting the timed trials for each \textit{Display Window}.
After finishing the trials for each Display Window, we asked them to complete a NASA TLX and VRSQ questionnaire.
The study session lasted approximately 90 minutes for each participant.

\subsection{Measurements}
We record trial time -- the duration from initiating movement by pressing the trigger button until the target is correctly placed within the frame boundary and the trigger button is pressed (and released) to confirm the selection. 
We logged additional attempts if the target was outside the frame at selection. 
During a trial, we measured total head rotation as the change in horizontal (yaw) rotation. 
The number of times the target entered and exited the center frame was recorded as the number of crossings. 
To assess perceived workload using the \textit{Mapping Functions}, participants completed the NASA Task Load Index (NASA TLX) questionnaire~\cite{nasatlx}. 
For perceived VR sickness, they filled out the Virtual Reality Sickness Questionnaire (VRSQ) questionnaire~\cite{kim2018virtual, jeon2020factors}.

\subsection{Results}

We analyzed trial time, total head rotations, and number of crossings with repeated measures ANOVAs and pairwise comparisons with Bonferroni corrections. 
We analyzed the additional attempts, NASA-TLX, and VRSQ data with Friedman tests followed by the Wilcoxon tests for pairwise comparisons. 


\subsubsection{Trial Time} 
We found that \textit{Mapping Function} had a significant effect on trial time ($F_{6, 93.49} = 106.69$, $p < 0.001$, $\eta^2$ = 0.80) (see Figure \ref{fig:study1_a}a).
Pairwise comparisons revealed that participants were significantly faster with polynomial (Mean M: 3.70s, Confidence Interval, CI: [lower-bound 3.45, upper-bound 3.96]), sigmoid (M: 3.94s, CI: [3.65, 4.24]) and linear (M: 4.36s, CI: [4.08, 4.63]) were significantly faster (all $p<0.001$) than with additive (M: 6.16s, CI: [5.71, 6.61]), continuous (M: 6.09s, CI: [5.77, 6.42]), friction (M: 6.10s, CI: [5.83, 6.36]), and interrupted (M: 6.54s, CI: [6.05, 7.03]).  
Additionally,  only the polynomial mapping was significantly faster ($p<0.001$) than the linear mapping.
We did not observe any other statistically significant differences.

We observed significant main effects for \textit{Target Distance} ($F_{2, 54} = 1224.22$, $p < 0.001$, $\eta^2$ = 0.98) (see Figure \ref{fig:study1_a}a).
The mean trial time is 4.24s (CI: [3.99, 4.47]) for \textit{Target Distance} 500cm, 5.17s (CI: [4.91, 5.44]) for 750cm, and 6.40s (CI: [6.10, 6.70]) for 1000cm.
Pairwise comparisons revealed that trials for the 500cm \textit{Target Distance} are significantly faster than trials at distances 750cm and 1000cm (all $p<.001$). 
In addition, trials at 750cm \textit{Target Distance} are significantly faster than 1000cm ($p<.0001$). 
We did not find a significant effect of \textit{Display Window} ($F_{2, 54} = 2.48$, $p = 0.09$) on trial time.

We found significant interactions of \textit{Mapping Function} $\times$ \textit{Target Distance} ($F_{12, 156} = 12.56$, $p < 0.001$, $\eta^2$ = 0.49) and  \textit{Mapping Function} $\times$ \textit{Display Window} ($F_{12, 156} = 12.56$, $p < 0.001$, $\eta^2$ = 0.49). 
For \textit{Mapping Function} $\times$ \textit{Target Distance}, we found that polynomial was significantly faster ($p<0.001$) than linear at the 750cm (M: 3.61s, CI: [3.33, 3.88]) vs. (M: 4.24s, CI: [3.96, 4.52]), and 1000cm (M: 4.40s, CI: [4.09, 4.71]) vs. (M: 5.47s, CI: [5.07, 5.87]) distances, though not at 500cm. The speed advantage of the sigmoid mapping (M: 4.57s, CI: [4.19, 4.94]) was significant compared to the linear mapping, but only for the 1000cm distance ($p<0.01$).
For \textit{Mapping Function} $\times$ \textit{Display Window}, the polynomial and sigmoid mappings became significantly faster as the window size increased. For the polynomial, trial time decreased from a mean of 3.86s (CI: [3.60, 4.12]) with the 400cm window to 3.59s (CI: [3.31, 3.86]) with the 800cm window ($p < 0.01$). 
For sigmoid, trial time decreased from a mean of 4.18s (CI: [3.85, 4.51]) at 400cm to 3.82s (CI: [3.52, 4.12]) at 800cm ($p < 0.001$). 
In contrast, the interrupted technique became significantly slower ($p < 0.05$) with the 800cm window (M: 6.82s, CI: [6.19, 7.45]) compared to the 600cm window (M: 6.31s, CI: [5.86, 6.76]).
We did not find any other significant interaction effects.

\begin{figure*}[!h]
    \centering
    \includegraphics[width=1.0\textwidth]{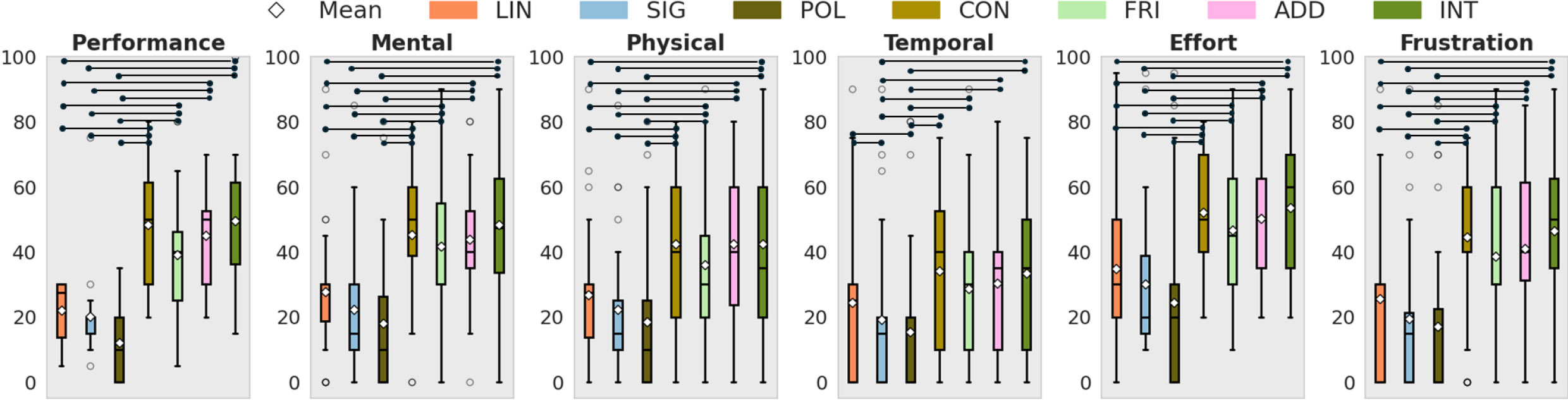}
	\caption{Subjective feedback score (NASA-TLX \cite{nasatlx}) for each \textit{Mapping Function}. Boxplots displays the median and IQR; whiskers denote 1.5$\times$IQR; circles indicate outliers; diamonds mark means. \update{Capped horizontal bars denote significance.}} 
    \Description{Figure 6 boxplots show subjective feedback scores for different mapping functions across various NASA-TLX workload metrics, including Mental Demand, Physical Demand, Temporal Demand, Performance, Effort, and Frustration. In the Performance metric, Polynomial, Sigmoid, and Linear mapping functions exhibit higher performance scores compared to Additive and Interrupted functions. For Mental Demand, Additive and Interrupted functions show higher demand, while Linear and Sigmoid show lower values. Physical Demand is similarly higher for Additive and Interrupted functions, with Polynomial and Linear displaying lower values. Temporal Demand shows relatively uniform scores across functions, but Additive and Interrupted generally appear to have higher values. Effort scores follow a similar trend, with Additive and Interrupted requiring more effort, while Polynomial and Linear demand less. Frustration levels are highest with Additive and Interrupted functions, with lower scores for Sigmoid and Linear. Overall, Polynomial, Sigmoid, and Linear mapping functions perform better across most metrics, showing lower demand and higher performance compared to Additive and Interrupted mappings.}
	\label{fig:study1_b} 
\end{figure*}

\subsubsection{Total Head Rotation} 
We found significant main effects for \textit{Mapping Function} ($F_{2.61, 70.66} = 17.74$, $p < 0.001$, $\eta^2$ = 0.40) (Figure \ref{fig:study1_a}b). Pairwise comparisons revealed that the sigmoid (M: 101.55\textdegree, CI: [87.47\textdegree, 115.63\textdegree]), polynomial (M: 110.21\textdegree, CI: [96.81\textdegree, 123.62\textdegree]) and linear (M: 113.28\textdegree, CI: [100.64\textdegree, 125.93\textdegree]) leads to significantly lower head rotations (all $p<0.01$) than additive (M: 191.29\textdegree, CI: [159.47\textdegree, 223.11\textdegree]), Continuous (M: 169.99\textdegree, CI: [148.34\textdegree, 191.65\textdegree]), and Friction (M: 161.61, CI: [140.80, 182.42]). 
Additionally, interrupt (M: 145.19\textdegree, CI: [116.72\textdegree, 173.11\textdegree]) had significantly lower head rotations than additive.

We observed significant main effects for \textit{Target Distance} ($F_{2, 54} = 79.37$, $p < 0.001$, $\eta^2$ = 0.75) on total head rotation (see Figure \ref{fig:study1_a}b). 
The mean head rotation is 129.49\textdegree~(CI: [116.26\textdegree, 142.72\textdegree]) for \textit{Target Distance} 500cm, 140.71\textdegree~(CI: [125.39\textdegree, 156.03\textdegree]) for 750cm, and 155.42\textdegree~(CI: [138.50\textdegree, 172.34\textdegree]) for 1000cm.
Post-hoc pairwise comparisons revealed that target at a \textit{Target Distance} of 500cm requires significantly less head rotation than targets at 750cm and 1000cm (all $p<.001$). 
In addition, targets with \textit{Target Distance} 750cm require significantly less head rotation than targets at 1000cm ($p<.0001$). 

We found a significant effect of \textit{Display Window} ($F_{1.63, 44.06} = 11.81$, $p <0.001$) on total head rotation.  
A \textit{Display Window} of 800cm (M: 152.51\textdegree, CI: [133.53\textdegree, 171.49\textdegree]) lead to significantly higher (all $p$'s $< .01$) head rotation than 400cm (137.80\textdegree, CI: [123.43\textdegree, 152.18\textdegree]) and 600cm (M: 135.31\textdegree, CI: [122.16\textdegree, 148.45\textdegree]).
We also observed significant interactions of \textit{Mapping Function} $\times$ \textit{Target Distance} ($F_{6.01, 162.23} = 10.72$, $p < 0.001$, $\eta^2$ = 0.29) and \textit{Mapping Function} $\times$ \textit{Display Window} ($F_{5.12, 138.43} = 2.43$, $p < 0.05$, $\eta^2$ = 0.08). 
For \textit{Mapping Function} $\times$ \textit{Display Window} showed that the interrupted technique required significantly more ($p<0.01$) rotation in the 800cm \textit{Display Window} (M: 167.74\textdegree, CI: [131.14, 204.34]) compared to the 400cm (M: 132.67\textdegree, CI: [108.69, 156.65]), and 600cm (M: 135.15\textdegree, CI: [106.21, 164.09]).

For \textit{Mapping Function} $\times$ \textit{Target Distance}, we found that increasing the distance significantly increased total head rotation for all mapping functions except sigmoid.
 While other techniques required significantly more head rotation to acquire farther targets, there was no significant difference in total head rotation for sigmoid across the 500cm (M: 100.90\textdegree, CI: [86.06, 115.72]), 750cm (M: 102.52\textdegree, CI: [87.87, 117.17]), and 1000cm (M: 101.24\textdegree, CI: [87.13, 115.35]) \textit{Target Distance} (all pairwise $p=1.000$).
No other interaction effect was found.


\subsubsection{Number of Crossings} 
We found significant main effects for \textit{Mapping Function} ($F_{3.97, 107.21} = 14.37$, $p < 0.001$, $\eta^2$ = 0.35) on the number of crossings (see Figure \ref{fig:study1_a}c). 
Pairwise comparisons showed that interrupted (M: 0.23, CI: [0.17, 0.29]), continuous (M: 0.25, CI: [0.20, 0.30]), friction (M: 0.25, CI: [0.22, 0.29]) and linear (M: 0.29, CI: [0.21, 0.36]) have significantly fewer crossings than additive (M: 0.46, CI: [0.39, 0.53]) and sigmoid (M: 0.45, CI: [0.35, 0.55]).
Additionally, the interrupted has significantly fewer crossings than the polynomial (M: 0.36, CI: [0.28, 0.43]).

We found significant main effects for \textit{Display Window} ($F_{2, 54} = 3.48$, $p < 0.05$, $\eta^2$ = 0.24).
Mean number of crossings for each size was: 400cm (M: 0.35, CI: [0.29, 0.40]), 600cm (M: 0.31, CI: [0.26, 0.35]), and 800cm (M: 0.32, CI: [0.27, 0.38]).
We found the medium size, 600cm, had significantly fewer crossings compared to the smaller 400cm display window.
This indicates that, as the display window size increases, the number of crossings decreases.
We did not find a significant effect of \textit{Target Distance} ($F_{2, 54} = 3.07$, $p = 0.054$) on number of crossings. 
The mean number of crossings for each \textit{target distance} was: 500cm (M: 0.34, CI: [0.29, 0.40]), 750cm (M: 0.32, CI: [0.27, 0.37]), and 1000cm (M: 0.32, CI: [0.27, 0.37]).

We observed a significant interaction effect of \textit{Mapping Function} $\times$ \textit{Display Window} ($F_{12,324} = 1.90$, $p < 0.05$, $\eta^2 = 0.07$). For \textit{Mapping Function} $\times$ \textit{Display Window}, we found significant difference only for the Sigmoid, where number of crossing was significantly higher ($p < 0.05$) with the 400cm \textit{Display Window} (M: 0.55, CI: [0.40, 0.70]) compared to both the 600cm (M: 0.41, CI: [0.31, 0.50]) and the 800cm \textit{Display Window} (M: 0.40, CI: [0.30, 0.49]) sizes. 
We did not find any other interaction effect.

\subsubsection{Additional Attempts} 
We observed significant effects for \textit{Mapping Function} ($\chi^2(6, N = 28) = 13.88$, $p < 0.05$) on additional attempts. 
Post-hoc pairwise comparisons among \textit{Mapping Function} ($\alpha$-level from 0.05 to 0.0024) found no significant differences between the pairs.
Median (IQR) additional attempts for each \textit{Mapping Function}: linear (median 0.08, IQR: 0.04 to 0.18), sigmoid (median 0.09, IQR: 0.03 to 0.15), friction  (median 0.10, IQR: 0.04 to 11.32), continuous (median 0.12, IQR: 0.04 to 0.18), polynomial (median 0.14, IQR: 0.06 to 0.19), interrupted (median 0.05, IQR: 0.14 to 0.20), and additive (median 0.06\%, IQR: 0.17 to 0.24). 
We found that linear and sigmoid had significantly fewer additional attempts than additive. 
We did not observe a significant effect of \textit{Target Distance} ($\chi^2(2, N = 28) = 0.50$, $p=0.78$) or \textit{Display Window} ($\chi^2(2, N = 28) = 2.79$, $p=0.25$) on additional attempts.
The median number of additional marker attempts for each \textit{Target Distance} was 0.13 (IQR: 0.04 to 0.19) 500cm, 750cm 0.13 (IQR: 0.07 to 0.18) for 750cm, and 0.12 (IQR: 0.05 to 0.21) for 1000cm. 
The median number of additional marker attempts for each \textit{Display Window} were 0.13 (IQR: 0.06 to 0.18) for 400cm, 0.13 (IQR: 0.05 to 0.20) for 600cm, and 0.13 (IQR: 0.04 to 0.15) for  800cm.

\subsubsection{VRSQ}
We found significantly different Oculomotor scores ($\chi^2 (6,N=28)=89.20, p<0.001$), Disorientation scores ($\chi^2 (6,N=28)=80.88, p<0.001$), and the total VRSQ scores ($\chi^2 (6,N=28)=86.47, p<0.001$). 
Post-hoc pairwise comparisons (Bonferroni: $\alpha$-levels from 0.05 to 0.0024) reveal that polynomial, sigmoid and linear have significantly lower Oculomotor, Disorientation and VRSQ-total scores than additive, interrupted, friction, and continuous. 
This indicates that additive, interrupted, friction, and continuous techniques caused a stronger sense of VR sickness in participants compared to the other techniques.

\subsubsection{NASA TLX}
We found significant differences for all NASA TLX metrics (Figure \ref{fig:study1_b}): Performance ($\chi^2 (6,N=28)=116.14, p<0.001$), Mental Demand ($\chi^2 (6,N=28)=93.18, p<0.001$), Physical Demand ($\chi^2 (6,N=28)=86.00, p<0.001$), Temporal Demand ($\chi^2 (6,N=28)=58.43, p<0.001$), Effort ($\chi^2 (6,N=28)=74.33, p<0.001$),  Frustration ($\chi^2 (6,N=28)=84.83, p<0.001$), and Overall task load ($\chi^2 (6,N=28)=104.18, p<0.001$). 
Post-hoc pairwise comparisons (Bonferroni: $\alpha$-levels from 0.05 to 0.0024) reveal that
polynomial, sigmoid, and linear had significantly higher performance, lower mental demand, lower physical demand, lower effort, lower frustration, and better overall ratings than additive, continuous, friction, and interrupted.
Additionally, polynomial and sigmoid outperformed continuous, friction, interrupted, and linear regarding temporal demand.
 Friction had significantly higher performance than Continuous. 
 The complete results of all pairwise comparisons are available in the supplementary material.

\subsubsection{Mapping Function Ranking}
Twenty-two out of our 28 participants ranked the polynomial function as their top choice for workspace movement. Three participants favoured the linear function, two selected the sigmoid, and one participant chose the friction mapping as their first choice. 
Conversely, 15 participants ranked the interrupted function as the least preferred (i.e., in 7th place). Six participants considered the continuous function, five the additive, and two the friction function as the least preferred.

\subsection{Answering RQs}

\paragraph{\textbf{RQ1: Effective mapping function}} 
Results show that polynomial, sigmoid and linear mappings outperformed continuous, friction, interrupted, and additive in trial time. Polynomial was also significantly faster than linear.  Polynomial, sigmoid, and linear required less head rotation, allowing participants to rotate the workspace with little head movement. The additive mapping, being difficult to control, showed the highest number of crossings, and sigmoid also produced more crossings than most other mappings, whereas interrupted yielded the fewest. According to NASA-TLX ratings, the polynomial mapping function had the lowest perceived workload. The workload for sigmoid and linear is also low; polynomial and sigmoid outperformed linear in temporal demand. For simulator sickness (VRSQ), polynomial, sigmoid, and linear mappings had significantly lower scores than continuous, friction, interrupted, and additive functions. Polynomial mapping was preferred by 22/28 participants (78.6\%), with linear favored by three participants (10.7\%), sigmoid by two (7.1\%), and friction by one (3.6\%). In line with previous research \cite{fashimpaur2023investigating, tsandilas2013modeless}, we conclude that the polynomial mapping is the most effective choice for head-based workspace navigation on large curved displays, due to its fastest trial times, highest subjective ratings.

\paragraph{\textbf{RQ2. Impact of display window size}}
In our study, participants' task was to bring off-screen targets to the center of the display window. 
Similar to prior work \cite{ens2016moving, sultana2024exploring}, our study did not find a significant effect of display window size on trial time or the number of additional attempts. This suggests that users' performance (in terms of trial time and additional attempts) when accessing off-screen information with head movement does not depend on the display window size. Instead, it is more influenced by the mapping function. 

\paragraph{\textbf{RQ3. Effect of target distance on navigation efficiency}}
The target distance had a significant impact on both trial time and the total head rotation. As expected, greater distances resulted in longer trial times -- also observed in prior studies \cite{hinckley2002quantitative, cao2008peephole, kopper2010human, ens2016moving, kaufmann2012revisiting}.
Larger target distances resulted in more head rotation than with shorter distances. We also observed that this effect was consistent across all mapping functions. However, the target distance did not significantly affect the number of additional attempts or the number of crossings. Accordingly, we conclude that the target distance significantly affects navigation efficiency in terms of speed and physical effort, but does not compromise user accuracy.
\section{Study 2 -- Navigation techniques for a map navigation task}
Our second study compares the performance of our head movement-based workspace navigation technique, using the best-performing mapping function from Study 1 -- polynomial rate control (which we refer to as ``\textit{Head-Polynomial}'' from here) -- against two industry-standard techniques: controller-based \drag and joystick-based \push. 
\update{In Study 1, we focused on users' performance of a set of head-based mapping functions to bring off-screen content to on-screen under controlled conditions, commonly used to evaluate techniques for accessing out-of-view targets \cite{radle2014bigger, santos2018exploring}, providing a clear comparison of the mapping functions. Building on these results, in Study 2, we evaluate the best-performing technique against industry-standard solutions in an ecologically valid task, capturing usability, user preferences, and relevant contextual factors. By including both controlled evaluation (Study 1) and realistic scenarios (Study 2), the two studies provide a comprehensive understanding of both fundamental performance and real-world applicability. This approach ensures that our findings are both scientifically rigorous and practically relevant.}

Our Study 2 has one research question:

\begin{itemize}
    \item \textbf{RQ4}) How does the \head technique compare to the baseline \drag and \push in terms of trial time, task accuracy, and perceived workload during map navigation tasks on a large curved display?
    
\end{itemize}

\subsection{Apparatus}
We used a similar study setup to that used in Study 1. 
We used the same curved display and the OptiTrack system to capture participants' head, hand, and controller movements. 
We used Unity 3D (with C\#) to develop the study application.

\subsection{Navigation Techniques}
We evaluated three techniques, \drag, Push--\&--Release, and Head-Polynomial, as outlined below.

\subsubsection{Drag--\&--Flick}
The \drag technique is a controller-based method that uses the familiar `click-and-drag' metaphor commonly employed with a mouse for navigating maps on desktop systems.
With ``Drag'', the user positions the cursor on the workspace and then holds the controller’s button to drag the map in correspondence with the cursor’s movement. 
The technique maintains a 1:1 mapping between the angular displacement of the cursor and the workspace.
For a large workspace, users may need to repeat this action multiple times -- a process known as `clutching' --  which involves initiating the drag, moving the map, releasing, repositioning, and repeating.
Researchers showed that such repetitive actions can reduce efficiency \cite{avery2014pinch}.
To mitigate extensive clutching, the user can also perform a ``Flick'' by leveraging the controller's inertial momentum, which quickly navigates through the map with a rapid flick motion of the controller. 
We use the following formula to perform drag and flick for this technique:

The normalized workspace velocity, $y$, is determined by the following piecewise function:
\begin{equation}
y = 
\begin{cases} 
G \cdot x & \text{Drag (button held)} \\
\text{sign}(x_f) \cdot \max(0, |M_f \cdot x_f| - D_f \cdot t_f) & \text{Flick (after release)}
\end{cases}
\label{eqn:drag_flick}
\end{equation}
where $y$ is the normalized workspace velocity; $x$ is the normalized instantaneous velocity of the controller; and $G$ is a gain factor. During the \textit{Flick} phase, the initial velocity is based on the controller's normalized velocity at the moment of release, $x_f$, amplified by a flick multiplier, $M_f$. This velocity then decays at a rate determined by the damping factor, $D_f$, over the time elapsed since release, $t_f$.

\begin{figure*}[!h]
    \centering
    \includegraphics[width=1.0\textwidth]{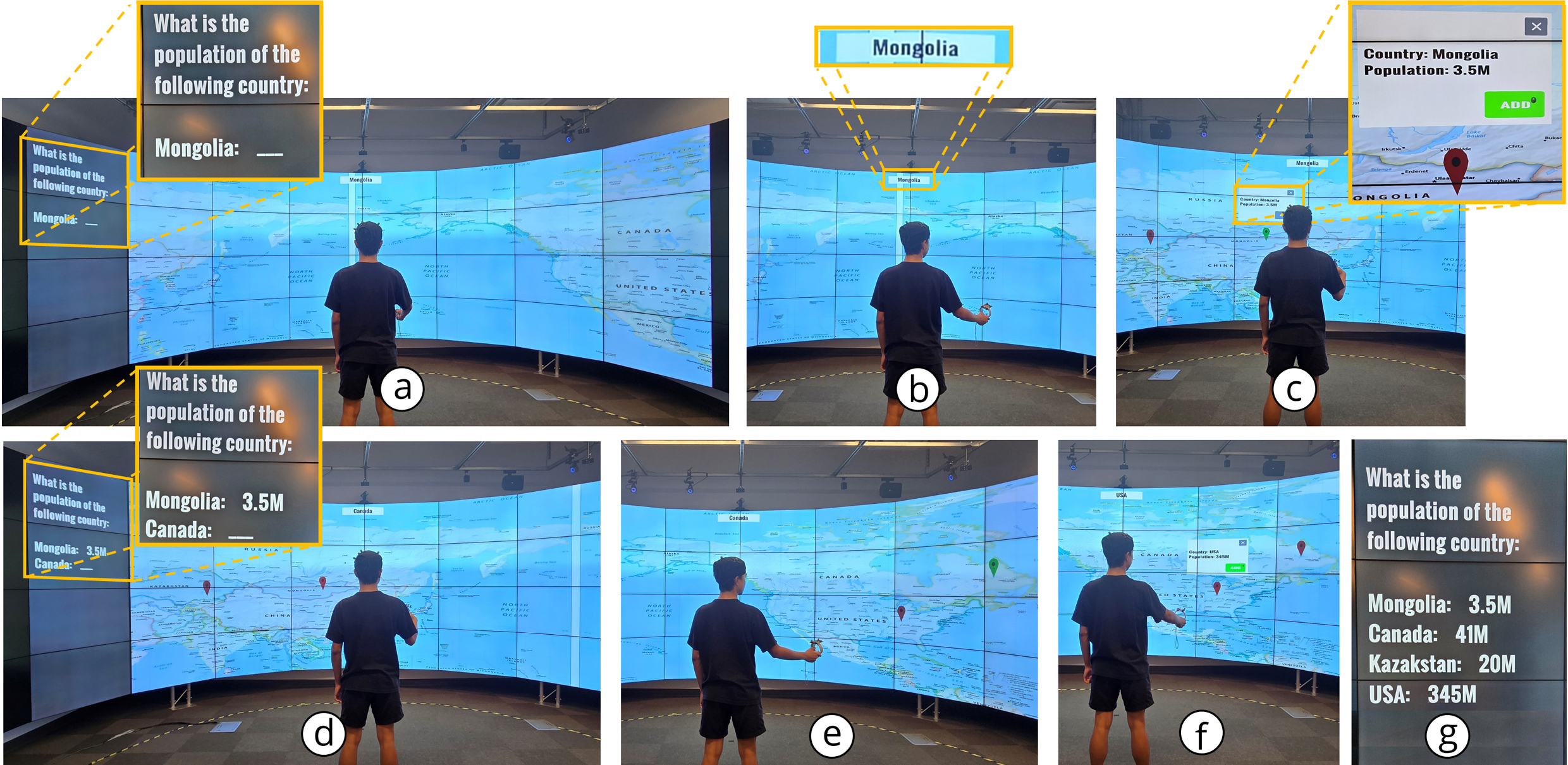}
  \caption{The experimental task flow for a single trial. (a) A trial begins with a prompt on the left panel asking for the population of a target country (e.g., Mongolia). (b) The country name always appears on the screen as a reminder. The participant navigates the map to find the target country, which is initially off-screen. (c) Once the country is on-screen, the user clicks the country's marker to open an information window and presses `ADD' button to add it as the answer to the question. (d) The question panel updates with the added information and prompts for the next country. (e-f) This search-and-select process is repeated for all four countries in the set. (g) The trial concludes once all four populations have been successfully added to the panel.}
    \Description{Figure 8 is a 7-panel figure showing the sequence of a user study task. Panel A shows a user facing a large curved screen with a question on the left: "What is the population of Mongolia?". Panel B shows the user navigating the map. Panel C shows the user selecting a pin on Mongolia, which opens a pop-up with its population. Panel D shows the question panel updated with Mongolia's population, now asking for Canada's. Panels E and F show the user continuing to navigate. Panel G shows the completed task, with the populations of four countries listed on the left panel.}

	\label{fig:s2-task} 
\end{figure*}

\subsubsection{Push--\&--Release}
With Push--\&--Release, the user pushes the controller joystick left or right to move the workspace in the corresponding direction.
If the joystick stays still, the map does not move.
Once the user pushes the joystick, the Meta Quest 2 controller~\cite{meta} provides a displacement value between -1 and 1, representing the extent to which the user pushes the joystick left or right. 
This value is then passed to the polynomial function (Equation \ref{eqn:polynomial}) used in \head to generate the speed at which the map is moved.
We ensured identical parameter settings to the Push--\&--Release technique as in the head-based polynomial mapping, using $b=2$ as the exponent and capping the maximum workspace velocity at $y=1$ (100 deg/sec).
\push allows users to navigate long distances by simply pressing the joystick, making it efficient for extended traversals without repeated clutching \cite{sargunam2018evaluating}.
However, prior work pointed out that holding a controller breaks the illusion of natural direct manipulation \cite{reski2020open}.
Since \push is the standard way to navigate workspaces on most curved display interfaces \cite{mechdyne-new1}, we included it as a baseline method.

\subsubsection{Head-Polynomial} 
We use the \head technique from Study 1 (see Figure \ref{fig:rate_control}d), which allows users to move the workspace by leveraging head movements while pressing and holding the controller button.
In this study, we also use the user’s horizontal head rotation (i.e., yaw) to control workspace movement.
We use the formula in Equation~\ref{eqn:polynomial} to calculate the movement speed of the map with respect to the head movement of the user.

\subsection{Participants} We recruited 18 participants (14 males and 4 females) with an average age of 24.78 years (SD = 3.28). Participants were compensated \$15 for their participation. Thirteen participants had experience using a wall-sized display (e.g., museum displays) within the last year, whereas five had never used one before.

\begin{figure*}[!h]
    \centering
    \includegraphics[width=1.0\textwidth]{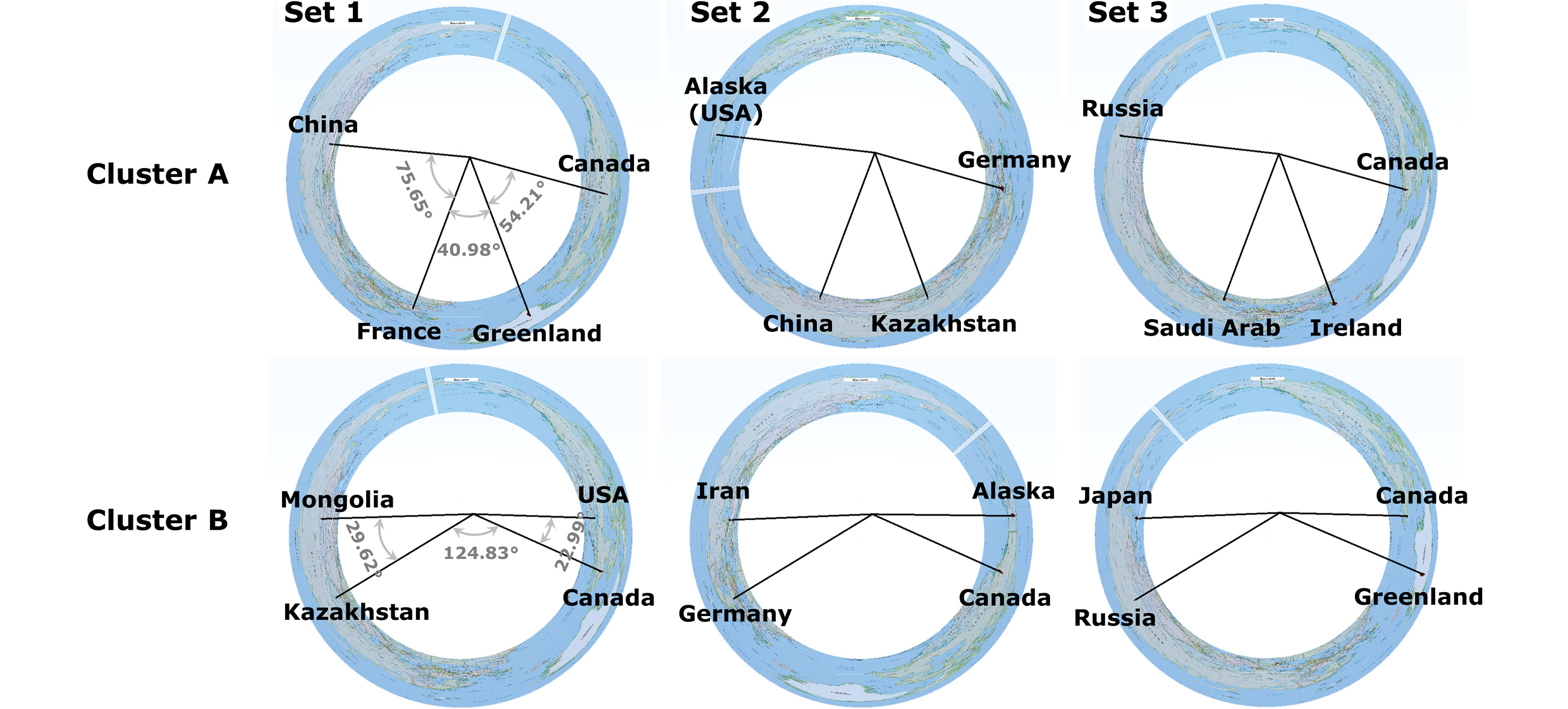}
    \caption{The two country clusters used in the Study 2. \textbf{Cluster A} features four countries with angular separations of 75.65\textdegree, 40.98\textdegree, and 54.21\textdegree). \textbf{Cluster B} presents a more varied spatial layout with separations of 29.62\textdegree, 124.83\textdegree, and 22.99\textdegree. For each cluster, we generated three distinct country sets by rotating the map while preserving the relative inter-marker distances.}
    \Description{Figure 7 displays a grid of six diagrams illustrating two country clusters, labeled Cluster A and Cluster B, each presented in three rotated configurations labeled Set 1, Set 2, and Set 3. The top row is dedicated to Cluster A, which maintains fixed angular separations of 75.65, 40.98, and 54.21 degrees between four countries; the countries shown are China, Canada, Greenland, and France in Set 1; Alaska (USA), Germany, Kazakhstan, and China in Set 2; and Russia, Canada, Ireland, and Saudi Arab in Set 3. The bottom row shows Cluster B, which has varied angular separations of 29.62, 124.83, and 22.99 degrees; its sets include USA, Canada, Kazakhstan, and Mongolia in Set 1; Alaska, Canada, Germany, and Iran in Set 2; and Japan, Canada, Greenland, and Russia in Set 3.}
    \label{fig:s2-country}
 \end{figure*}

\subsection{Task and Procedure}
The study task required participants to locate and select four countries on a 360\textdegree~world map and gather information from markers located in the countries. 
To minimize spatial familiarity, we created two country clusters, Cluster A and Cluster B.
For Cluster A, the angular separations between adjacent country markers are 75.65\textdegree, 40.98\textdegree, and 54.21\textdegree; for Cluster B, they are 29.62\textdegree, 124.83\textdegree, and 22.99\textdegree~, as shown in Figure \ref{fig:s2-country}. 
We then use the angles and rotate the map to generate three country sets under each cluster: Set 1, Set 2, and Set 3.
This step preserves the relative inter-marker distances across the four countries while generating three unique country sets within each country cluster.
Each country in a set is visually identified with a red map marker, see Figure \ref{fig:s2-task}c-f.
Once the participant clicks on a marker, a small window opens, displaying the country name, the population, and an `add' button, as shown in Figure~\ref{fig:s2-task}c.

At the start of each trial, a question, i.e., \textit{What is the population of the following countries?}'', is prompted in a 100cm wide panel located at the left side of the display, see Figure~\ref{fig:s2-task}a.
Below the question, the name of a country is displayed.
The map is positioned to the right of the panel. The display area for the map is 700cm wide. 
The country name displayed in the side panel is also shown in the middle of the display, serving as a constant reminder within sight, cf. Figure~\ref{fig:s2-task}b).
Initially, the map is displayed in a way so that no countries from the country set are visible on the display.
The participant is first required to use one of the techniques to initiate map movement and bring the prompted country to the display to answer the question.
\begin{figure*}[t]
    \centering
   \includegraphics[width=1.0\textwidth]{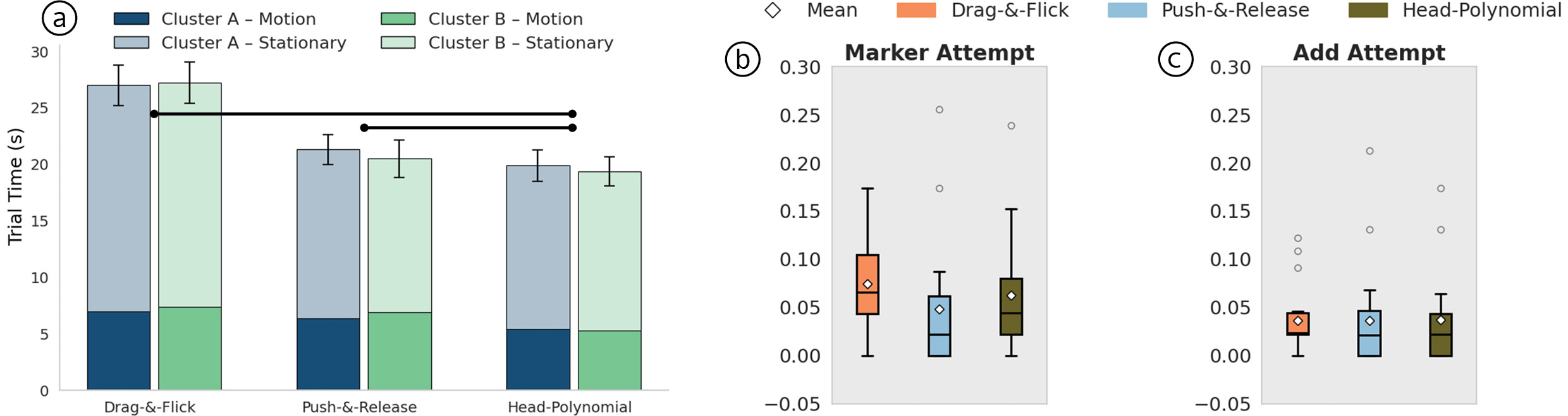}
	\caption{
    Study~2 results. (a) Trial time by \textit{Navigation Technique}, partitioned into Motion and Stationary time for each \textit{Country Cluster}. Error bars show 95\% confidence intervals \update{and capped horizontal bars denote significance.} (b) Distribution of wrong marker attempts per trial by technique. (c) Distribution of wrong add attempts per trial by technique. In (b)–(c), boxplots show median and IQR, whiskers denote 1.5$\times$IQR, and diamonds mark means. } 
    \Description{Figure 9 presents the results of Study 2 in three panels, labeled (a), (b), and (c).
Panel (a) is a stacked bar chart showing the total trial time for three navigation techniques: Drag-&-Flick, Push-&-Release, and Head-Polynomial. Each bar is divided into a darker bottom portion representing "Motion time" and a lighter top portion for "Stationary time." Results are shown for two conditions: Country Cluster A (in shades of blue) and Country Cluster B (in shades of green). The chart indicates that the Head-Polynomial technique was the fastest overall, with the lowest combined motion and stationary time. The Drag-&-Flick technique was the slowest.
Panel (b) is a box plot showing the distribution of wrong "Marker Attempts" per trial for the three navigation techniques. The techniques are color-coded: Push-&-Release is light blue, Drag-&-Flick is orange, and Head-Polynomial is olive green. The data shows that all three techniques resulted in a similarly low number of wrong marker attempts, with medians near or below 0.10.
Panel (c) is a box plot, using the same color scheme as panel (b), showing the distribution of wrong "Add Attempts" per trial. The results are similar to the marker attempts, indicating that all three navigation techniques led to a very low and comparable number of wrong add attempts, with medians close to 0.02.}
	\label{fig:study2_time_error} 
\end{figure*} 

Once the country is visible, the participant can click the marker to open the information window. With a click on the green ``Add'' button, the information is added to the question panel to the left of the display area, cf. Figure~\ref{fig:s2-task}c.
Once the information from the correct country is added, the system plays a sound to indicate a correct selection, which shows the next country in the panel, see Figure~\ref{fig:s2-task}d. 
If the participant opens an incorrect country window, no error notification is given. 
The participant is only notified with a sound signal when attempting to add information (i.e., when clicking the add button) for an incorrect country.
Participants cannot proceed to the next country until they have selected the correct country. 
Participants had to repeat the same process to select all four countries for a given trial correctly (Figure~\ref{fig:s2-task}e-f).
Once all countries were added (Figure \ref{fig:s2-task}g), the trial ended, and the question panel was updated with a new country. The map was then initialized to display no markers.


\subsection{Design}

We used a within‐subjects design with two independent variables: 1) \textit{Navigation Technique} with three levels, \head, \drag, and \push, and 2) \textit{Country Cluster} with two levels, Cluster A and Cluster B. 
The order of the \textit{Navigation Technique} $\times$ \textit{Country Cluster} combinations was counterbalanced across participants using a Latin square design. For each \textit{Navigation Technique} $\times$ \textit{Country Cluster} combination, we randomly selected one of the six country sets (from Figure \ref{fig:s2-country}). Once a set was assigned to a specific combination, it was not reused for others, following a non-repeated random selection procedure. This helped reduce the likelihood that participants could memorize country positions. For each combination of \textit{Navigation Technique} and \textit{Country Cluster}, participants performed 24 trials, where each trial corresponded to a different permutation of the four countries in the cluster (e.g., trial 1: find country in the order of A $\rightarrow$ B $\rightarrow$ C $\rightarrow$ D, trial 2: B $\rightarrow$ A $\rightarrow$ C $\rightarrow$ D and so on). 
Thus, the order of country selection varied across trials.
In total, each participants completed 3 \textit{Navigation Technique}  $\times$ 2 \textit{Country Cluster} $\times$ 24 trials = \updateNEW{144} trials. 
Prior to the main trial series for each \textit{Navigation Technique}, participants completed a set of practice trials to familiarize themselves with the technique. 
After completing each \textit{Navigation Technique}, participants filled out the NASA TLX and VRSQ questionnaires for that particular \textit{Navigation Technique}. 
A study session lasted approximately one hour.

\subsection{Measurement}
We define trial time as the duration from the initiation of workspace movement -- marked by pressing the controller button after the question appeared on the left panel -- to successful addition of the fourth country’s information to the panel.
The total motion time and total stationary time distinguish between periods when the workspace was in motion and when it was stationary.
We also recorded two error measures for each trial. Wrong ``Marker Attempt'' refers to the total number of times a participant clicked on a map marker for a country that was not the current target country, as indicated in the question panel. ``Add Attempts'' is the number of times a participant clicked the `Add' button in the pop-up window for an incorrect country.
Both measures are accumulated over the entire trial across all four countries and do not reset until participants advance to the next trial.
Example: If the panel first asks for Mongolia and the participant clicks the marker in Kazakhstan and presses `Add', the program records one Marker attempt and one Add Attempt.

\subsection{Results}
We analyzed trial times with a repeated measures ANOVA and pairwise comparisons with Bonferroni corrections. 
We used Friedman tests followed by Wilcoxon tests for pairwise comparisons to analyze the number of Marker Attempts and Add Attempts, NASA TLX, and VRSQ data. 

\begin{figure*}[ht]
   \centering
    \includegraphics[width=0.9\textwidth]{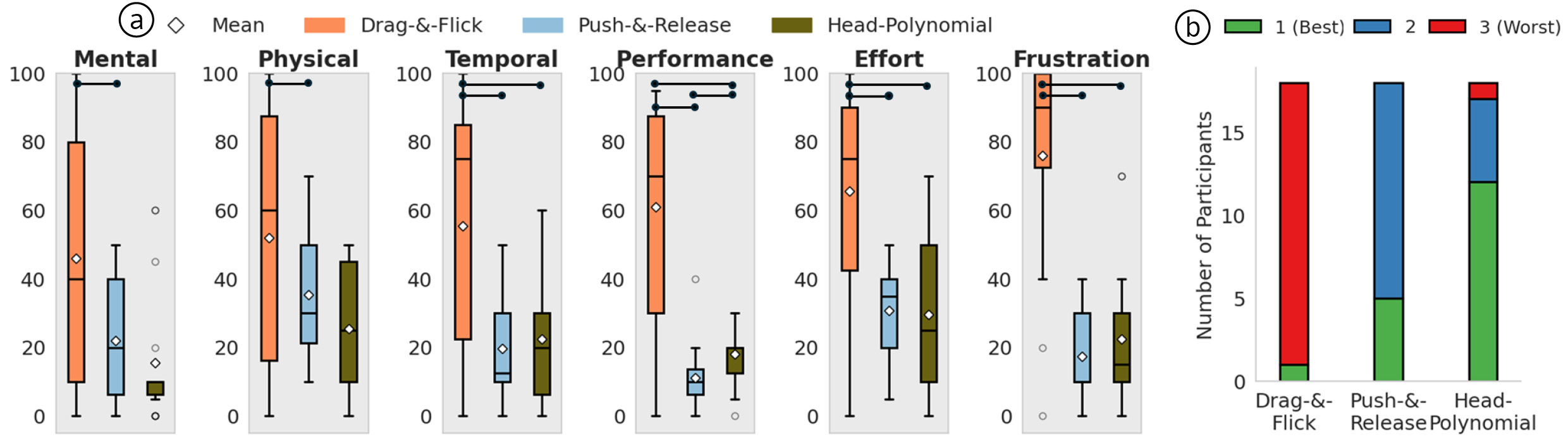}
	\caption{(a) NASA TLX ratings for Mental, Physical, Temporal, Performance, Effort, and Frustration across \textit{Navigation Techniques}. Boxplots show median and IQR; whiskers denote 1.5$\times$IQR; diamonds mark means. Lower scores indicate lower workload; for performance, a lower value indicates better perceived performance. \update{Capped horizontal bars denote significance.} (b) Participant preference rankings (counts of 1=\textit{best}, 2, 3=\textit{worst}) for each technique. } 
    \Description{Figure 10 presents subjective feedback and preferences for three navigation techniques in two panels, labeled (a) and (b).
Panel (a) consists of six box plots displaying NASA TLX workload ratings. Each plot corresponds to a specific metric: Mental, Physical, Temporal, Performance, Effort, and Frustration. Within each plot, three techniques are compared: Drag-&-Flick (orange), Push-&-Release (light blue), and Head-Polynomial (olive green). A consistent trend is visible across all six metrics: the Drag-&-Flick technique consistently received the highest (worst) workload scores, the Head-Polynomial technique received the lowest (best) scores, and the Push-&-Release technique's scores were in the middle.
Panel (b) is a stacked bar chart showing participant preference rankings for the three techniques. The height of each bar represents the total number of participants. The bars are segmented by color to show how many participants gave a rank of 1 (best, shown in green), 2 (shown in blue), or 3 (worst, shown in red). The chart shows that the Head-Polynomial technique was ranked as best by the majority of participants. The Push-&-Release technique was ranked as either best or second best. The Drag-&-Flick technique was ranked as worst by nearly all participants.}
\label{fig:study2_Nasatlx} 
\end{figure*}

\subsubsection{Trial Time}
We found that \textit{Navigation Technique} had a significant effect on trial time ($F_{2, 34} = 158.60$, $p < 0.001$, $\eta^2$ = 0.90), see Figure~\ref{fig:study2_time_error}a.
Pairwise comparisons revealed that \head (M: 19.57s, CI: [lower-bound 18.20, upper-bound 20.98]) is significantly faster than the other two techniques: (all $p<0.05$) \push (M: 20.87s, CI: [19.35, 22.39]) and \drag (M: 27.07s, CI: [25.24, 28.91]). 
We did not find a significant effect of \textit{Country Cluster} ($F_{1, 17} = 1.04$, $p = 0.32$) on trial time. 
We also did not find any interactions of \textit{Navigation Technique} $\times$ \textit{Country Cluster} ($F_{2, 34} = 1.58$, $p = 0.22$).


\textit{Motion Time}:
We found that \textit{Navigation Technique} had a significant effect on motion time ($F_{1.43, 24.38} = 6.54$, $p < 0.01$, $\eta^2$ = 0.28) (see Figure \ref{fig:study2_time_error}a).
Pairwise comparisons revealed that \head (M: 5.31s, CI: [lower-bound 4.63, upper-bound 6.00]) required significantly ($p<0.05$) lower motion time than both \push (M: 6.62s, CI: [5.85, 7.40]) and \drag (M: 7.14s, CI: [5.90, 8.39]). 
We did not find a significant effect of \textit{Country Cluster} ($F_{1, 17} = 1.81$, $p = 0.43$) on motion time. 
We also did not find any interaction of \textit{Navigation Technique} $\times$ \textit{Country Cluster} ($F_{2, 34} = 0.99$, $p = 0.38$).

\textit{Stationary Time}:
We found that \textit{Navigation Technique} had a significant effect on stationary time ($F_{2, 34} = 105.90$, $p < 0.001$, $\eta^2$ = 0.86), see Figure \ref{fig:study2_time_error}a.
Pairwise comparisons revealed that \head (M: 14.27s, CI: [lower-bound 13.28, upper-bound 15.27]) and \push (M: 20.87s, CI: [19.35, 22.39]) required significantly lower stationary time than \drag (M: 19.93s, CI: [18.54, 21.32]). 
However, we did not find a significant difference between the stationary time of \head and Push--\&--Release.
There were also no significant effect of \textit{Country Cluster} ($F_{1, 17} = 1.93$, $p = 0.18$) on stationary time or a significant \textit{Navigation Technique} $\times$ \textit{Country Cluster} interaction ($F_{2, 34} = 1.98$, $p = 0.15$).

\subsubsection{Additional Marker Attempts}
The overall mean number of additional attempts was 0.06.
We did not observe significant effects for \textit{Navigation Techniques} ($\chi^2(2, N = 18) = 4.90$, $p = 0.09$) on additional marker attempts (see Figure \ref{fig:study2_time_error}b). 
The median number of additional marker attempts for each of the three \textit{Navigation Techniques} was 0.06 (IQR: 0.04 to 0.11) for \drag , 0.04 (IQR: 0.02 to 0.09) for \head, and 0.02 (IQR: 0.00 to 0.07) for \push. 
The median number of additional marker attempts was 0.03 (IQR: 0.01 to 0.05) for Cluster A and 0.08 (IQR: 0.05 to 0.11) for Cluster B. This difference was significantly different ($\chi^2(2, N = 18) = 14.22$, $p < 0.001$).


\subsubsection{Additional Add Attempts}
The overall mean number of additional add attempts was 0.04.
We did not observe significant effects for \textit{Navigation Techniques} ($\chi^2(2, N = 18) = 0.22$, $p = 0.90$) on additional add attempts (see Figure \ref{fig:study2_time_error}c). 
The median number of additional add attempts was 0.02 (IQR: 0.02 to 0.05) for all three \textit{Navigation Techniques}.
For \textit{Country Cluster}, we found Cluster A (median 0.00, IQR: 0.01 to 0.03) had significantly lower ($\chi^2(2, N = 18) = 13.24$, $p < 0.001$) additional add attempts compared to Cluster B (median 0.03, IQR: 0.06 to 0.08).


\subsubsection{NASA TLX}
We used NASA TLX ratings to examine users’ perceived workload when using the Navigation techniques. 
We found significant differences for all workload metrics.
Table~\ref{tab:nasatlx_vrsq} lists the statistical results.
Post-hoc tests showed that \head consistently resulted in significantly lower workload scores than \drag for nearly all metrics.
\push also outperformed \drag in terms of performance, mental, physical, temporal, effort, frustration, and overall workload. \head outperformed \push for the performance metric, while no other differences between the two were significant. Overall, participants rated \head as the least demanding, followed by Push--\&--Release, with \drag being the most demanding. Figure~\ref{fig:study2_Nasatlx}a shows the workload ratings.

\subsubsection{VRSQ}
The results from VRSQ questions show significant effects for Oculomotor strain and Total VRSQ scores, while Disorientation showed no significant effect, as listed in Table~\ref{tab:nasatlx_vrsq}. 
Pairwise comparisons revealed that both \head and \push had significantly reduced oculomotor strain and overall sickness compared to Drag--\&--Flick. 
There were no significant differences between \head and Push--\&--Release. 
This indicates that \head offered the most comfortable experience, followed by Push--\&--Release, with \drag inducing the most symptoms. 

\subsubsection{Technique Ranking}
As displayed in Figure~\ref{fig:study2_Nasatlx}b, twelve out of 18 participants ranked \head as their top choice for navigation. Five participants preferred Push--\&--Release and only \updateNEW{one} preferred \drag. Conversely, 17 participants ranked \drag as the least preferred technique, while only one participant rated \head lowest. No participants ranked \push as their least preferred navigation technique.
\begin{table*}[htbp]
    \caption{Statistical results for NASA TLX and VRSQ scores across the \textit{Navigation Techniques}.}
    \Description{Table 1: 

Table 1 summarizes statistical test results comparing the three navigation techniques—Head-Polynomial, Push-&-Release, and Drag-&-Flick—on NASA TLX workload metrics and VRSQ cybersickness scores. The table is structured with five columns: the metric name, Friedman test results with a significance threshold of p<0.05, and three pairwise Wilcoxon signed-rank comparisons with Bonferroni-corrected thresholds of p<0.016: Head-Polynomial versus Drag-&-Flick, Push-&-Release versus Drag-&-Flick, and Push-&-Release versus Head-Polynomial.

For NASA TLX metrics, Friedman tests show significant overall effects for all six metrics: Performance (χ²(2)=25.04, p<0.001), Mental (χ²(2)=10.31, p<0.01), Physical (χ²(2)=10.90, p<0.005), Temporal (χ²(2)=19.74, p<0.001), Effort (χ²(2)=16.39, p<0.001), Frustration (χ²(2)=23.06, p<0.001), and Overall workload (χ²(2)=21.33, p<0.001). Pairwise tests indicate that both Head-Polynomial and Push-&-Release consistently perform significantly better than Drag-&-Flick for most metrics. For example, Performance comparisons show Z=-3.69 (p<0.001) for Head-Polynomial vs Drag-&-Flick and Z=-3.51 (p<0.001) for Push-&-Release vs Drag-&-Flick. Similarly, large significant differences are observed for Temporal, Effort, and Frustration. Comparisons between Head-Polynomial and Push-&-Release show no significant differences except for Performance (Z=-2.61, p<0.01), where Head-Polynomial outperforms Push-&-Release.

For VRSQ metrics, Oculomotor strain (χ²(2)=13.75, p<0.001) and Total score (χ²(2)=13.50, p<0.001) show significant overall differences, while Disorientation (χ²(2)=2.40, p=0.30) does not. Pairwise results show that both Head-Polynomial and Push-&-Release significantly reduce Oculomotor strain and Total VRSQ compared to Drag-&-Flick (all p<0.016), but no significant differences are found between Head-Polynomial and Push-&-Release.
}
    \centering
    \resizebox{\linewidth}{!}{

\setlength{\tabcolsep}{1pt}
\begin{tabular}{|c|c|c|c|c|}
 \hline
  & Friedman  & \head $\leftrightarrow$ \drag & \push $\leftrightarrow$ \drag & \push $\leftrightarrow$ \head \\
 & (Threshold: p$<$0.05) & (Threshold: p$<$0.016) & (Threshold: p$<$0.016) & (Threshold: p$<$0.016) \\
 \hline
 \multicolumn{5}{|c|}{\textbf{NASA-TLX}}\\
 \hline
  \textbf{Performance} & $\chi^2$(2)=25.04, p$<$0.001 & Z=-3.69, p$<$0.001 & Z=-3.51, p$<$0.001 & Z=-2.61, p$<$0.01 \\
 \hline
  \textbf{Mental} & $\chi^2$(2)=10.31, p$<$0.01 & Z=-2.35, p=0.019 & Z=-2.70, p$<$0.01 & Z=-1.82, p=0.069 \\
 \hline
  \textbf{Physical} & $\chi^2$(2)=10.90, p$<$0.005 & Z=-1.94, p=0.053 & Z=-2.68, p$<$0.01 & Z=-1.90, p=0.057 \\
 \hline
  \textbf{Temporal} & $\chi^2$(2)=19.74, p$<$0.001 & Z=-3.07, p$<$0.005 & Z=-3.32, p$<$0.001 & Z=-0.71, p=0.475 \\
 \hline
  \textbf{Effort} & $\chi^2$(2)=16.39, p$<$0.001 & Z=-2.86, p$<$0.005 & Z=-3.08, p$<$0.005 & Z=-0.31, p=0.754 \\
 \hline
  \textbf{Frustration} & $\chi^2$(2)=23.06, p$<$0.001 & Z=-3.66, p$<$0.001 & Z=-3.60, p$<$0.001 & Z=-1.28, p=0.199 \\
 \hline
  \textbf{Overall} & $\chi^2$(2)=21.33, p$<$0.001 & Z=-3.47, p$<$0.001 & Z=-3.33, p$<$0.001 & Z=-0.11, p=0.913 \\
 \hline
 \multicolumn{5}{|c|}{\textbf{VRSQ}}\\
 \hline
  \textbf{Oculomotor} & $\chi^2$(2)=13.75, p$<$0.001 & Z=-2.11, p=0.035 & Z=-2.56, p$<$0.016 & Z=-0.16, p=0.875 \\
 \hline
  \textbf{Disorientation} & $\chi^2$(2)=2.40, p=0.30 & Z=-1.69, p=0.091 & Z=-1.84, p=0.066 & Z=-0.06, p=0.953 \\
 \hline
  \textbf{Total} & $\chi^2$(2)=13.50, p$<$0.001 & Z=-2.11, p=0.035 & Z=-2.56, p$<$0.016 & Z=-0.23, p=0.820 \\
 \hline
\end{tabular}
}
    \label{tab:nasatlx_vrsq}
\end{table*}


\subsection{Answering RQ4}
Our RQ4 was \textit{How does the \head technique compare to the baseline \drag and \push in terms of trial time, task accuracy, and perceived workload during map navigation tasks on a large curved display?}

Our results demonstrate that the head-based navigation technique is an effective technique for navigating large curved displays, outperforming the two baselines. 
In terms of time, \head was significantly faster than both \push and Drag--\&--Flick. 
We attribute the superior performance of \head to several factors: (i) it enables continuous traversal without the need for clutching, (ii) it provides faster movement across large workspaces by leveraging natural head yaw, and (iii) it reduces physical effort compared to repetitive controller-based actions.
The \drag technique requires users to perform repetitive `clutching' actions to traverse the large workspace, a known issue that can reduce efficiency \cite{avery2014pinch}. 
In contrast, both \head and \push allow for continuous traversal over long distances with a single action (e.g., push joystick or rotate head), which is more efficient for this type of task \cite{sargunam2018evaluating}.

We found no significant effect of \textit{Navigation Technique} on either of the error criteria: Additional marker attempts and additional add attempts. The Additional attempts were low for all techniques, with medians close to zero for each technique. Thus, the speed benefit of \head did not trade off accuracy. Cluster A elicited fewer extra actions than Cluster B, which likely reflects its angular spacing used for the countries within the cluster. 

NASA-TLX results revealed that \head consistently produced lower workload scores than \drag across most subscales. Similarly, \push outperformed \drag on all subscales -- performance, mental, physical, temporal, effort, frustration, and overall workload. Between \head and \push, participants reported higher perceived performance with \head, while the scores for the other NASA-TLX subscales were comparable. Consistent with these findings, VRSQ results showed that both \head and \push yielded significantly lower oculomotor strain and overall VRSQ scores compared to \drag, indicating that continuous rate-controlled movement is more comfortable for users.

Finally, preference data strongly support these findings: 12 out of 18, ranked \head as their most preferred technique, while 17 out of 18 ranked \drag as their least preferred. The strong preference for \head, combined with its superior performance and comfort, highlights its potential as an effective navigation method for large curved displays.

\section{Discussion}
\subsection{External Validity and Practical Context}

\update{While our findings stem from tests in a laboratory setup, we envision that the proposed head-based navigation techniques can transfer to ``real-world'' environments with wall-sized curved displays. In recent years, we have seen an increased use of wall-sized curved displays. They have been adopted for immersive analytics and collaborative decision-making with complex datasets \cite{cavallo2019dataspace}, as well as for immersive training in flight, driving, and gaming \cite{pixelwix-curved}. Sectors relying on advanced visualization, such as aerospace and meteorology, are also deploying high-fidelity curved LED walls (e.g., the Sony Crystal LED system \cite{VisualDisplaysSony}) for collaborative and immersive reviews. Additionally, wall-sized curved surfaces are increasingly serving as digital canvases in museums and galleries for immersive digital art experiences \cite{strongmdi, snadisplays}}.

\update{Within these contexts, our results offer several transferable insights. Based on RQ1, we suggest that polynomial rate control is a promising mapping function for head-based scrolling on large displays, compared to linear or zone-control mappings. From RQ2, our findings indicate that performance is relatively robust across different display window sizes, suggesting flexibility in selecting display sizes. RQ3 highlights the cost of long target distances: larger angular separations increase task difficulty and thus frequently used information should be placed closer together, while less relevant content can be positioned further away. Finally, we recommend that designers consider using head movement as a complementary input alongside physical controllers, as we found from RQ4.}

\update{However, caution is warranted when generalizing our results to setups that differ substantially from our study configuration. Our studies were conducted on a curved display with a 3.27-meter radius, thus, setups with significantly more or less curvature may require recalibration of mapping parameters (e.g., the polynomial exponent) to maintain intuitive control. Furthermore, our participants stood at the center of curvature; in real-world settings, off-center positions may alter the angular sizes of windows and targets, potentially affecting selection time and accuracy -- an area for further investigation \cite{kopper2010human}. Finally, our study focused solely on horizontal (yaw) movements, as head turns left and right are the primary interaction axis for wide curved displays \cite{ragan2016amplified}. Applications requiring 2D navigation will need further evaluation to assess head-based control compared to controller-based input.}

\subsection{Design Guidelines}

We interpret the key insights from two studies and discuss the design implications to guide the development of future navigation techniques for large curved displays.

\subsubsection{Consider Rate Control with Polynomial Mapping}
Our findings from Study 1 show that, among the tested mapping functions, rate control functions outperformed zone control techniques in terms of trial time. Specifically, the polynomial mapping function emerged as the most effective choice, having the fastest trial times, the lowest perceived workload ratings, and the lowest simulator sickness scores. The polynomial function was also the favourite among participants, with nearly 70\% ranking it as their first choice. Furthermore, the polynomial function required less total head rotation compared to several other techniques, helping to mitigate potential fatigue. The success of the polynomial function in Study 1, along with its subsequent high performance in the realistic map task in Study 2, confirms its effectiveness. Accordingly, for head-based navigation on large displays, designers should consider using polynomial rate control as the core mapping function.

\subsubsection{Use Head-Based Technique for Workspace Navigation on Large Curved Display}
Study 2 showed that \head was significantly faster than the industry-standard joystick-based \push method and the Drag--\&--Flick. This speed advantage was achieved without any loss of accuracy, as no significant differences in errors were found across techniques. The \head technique also resulted in the lowest motion time, indicating it is a more direct and fluid way to navigate workspaces on large curved displays. Subjectively, participants rated their own performance highest with \head and ranked it as their most preferred option. Designers should therefore consider head-based navigation as a potential technique to enhance user efficiency and speed in tasks that require exploring large workspaces, freeing the user's hands for selection tasks rather than complex navigation.

\subsubsection{Avoid Clutching-Heavy \drag for Large Workspace Navigation}
The \drag technique performed poorly across nearly every metric in Study 2. 
It was the slowest method, largely due to the time users spent in a stationary state while performing repetitive `clutching' actions, known to reduce efficiency \cite{avery2014pinch}. 
Furthermore, \drag induced the highest perceived workload, frustration, and VR sickness scores, and was ranked as the least preferred technique by 17 out of 18 participants. 
Given these results, we recommend that designers avoid using click-and-drag metaphors as the default method for navigating large-scale environments.

\subsubsection{Offer Joystick-based \push as a Secondary Option}
While \push is not as fast as Head-Polynomial, the joystick-based technique proved to be a viable alternative to Drag--\&--Flick. 
It was faster than \drag and got low workload and VR sickness scores, comparable to those of Head-Polynomial. 
This suggests that joystick control is an efficient and less physically demanding method for users who may not prefer or are unable to use head-based navigation for extended periods. 
Therefore, we recommend that designers consider \push as a secondary option, particularly in scenarios where head mobility may be limited (e.g., accessibility needs, physically constrained environments).

\subsection{Limitations and Future Work}

Our study was conducted using a large curved display with a fixed viewing angle of 180\degree and a radius of 3.27 meters. This limits the generalizability of our findings to other display configurations. Future work could investigate whether our results are applicable to other configurations. 
\update{We did not have an equal number of female and male participants in our studies: Study 1: 20 male, 8 female; Study 2: 14 male, 4 female. This may limit the generalizability of our results across the sexes. This imbalance may also have influenced the subjective results regarding participants' experiences with head-based input (e.g., comfort, sickness, preferences).}
\update{Another key limitation of our studies is the short washout period for VR sickness in our studies: in both studies, participants tested all conditions within a single session -- following the approach used by other researchers \cite{xu2020results, hock2017carvr}. Kim et al. \cite{kim2018virtual}, who introduced VRSQ, employed this approach as well. 
This design may have led to confounding sickness data, as symptoms from one technique could carry over and affect the simulator sickness scores of a subsequent technique \cite{kasegawa2024effects}. Prior studies assessing VR/Simulator sickness include longer breaks (e.g., 24 hours \cite{clark2021rest}) between conditions to allow symptoms to fully subside. Future work should incorporate such a washout period to ensure a more accurate and independent measurement of sickness for each navigation technique.}

\update{Furthermore, while our use of a $\pm 10\degree$ stop zone and a long-press activation gesture helped mitigate the ``Midas Touch'' problem, this issue of unintended activation warrants further investigation. 
Accordingly, future work can explicitly study involuntary confirmations related to the Midas Touch problem for head-based workspace movement.
For instance, glancing past the $\pm $20\degree constant zone in our zone-control techniques could still trigger unwanted movement. 
Future work could explore alternative mitigation strategies such as requiring a brief dwell-time inside an activation zone, implementing ``soft zones'' that gradually ramp up control, or comparing different explicit activation gestures to find the optimal balance between responsiveness and precision.}
Future work could extend our one-dimensional navigation task to two- and three-dimensional navigation, allowing a more comprehensive evaluation of mapping functions in realistic 360\degree exploration scenarios.

It would also be interesting to explore how multiple users jointly navigate and interact with 360\degree content in co-located settings, including features such as synchronized viewpoints and awareness of shared head-rotation.
Future research could also investigate adaptive or personalized mapping functions that adjust sensitivity or control type based on user preferences or behavior. 
For instance, the system could switch between different techniques based on user performance and need, such as selecting a less demanding technique, such as \push when users are fatigued with Head-Polynomial.
Such approaches may improve performance, comfort, and overall user experience on large curved displays.
We do acknowledge that head-based navigation techniques might have an impact on users’ sense of presence and spatial awareness in large curved displays. 
Future studies could focus on examining how different mapping functions and techniques influence these factors and how they, in turn, affect users' task performance with off-screen content on large displays.

Lastly, our study focused on head movements for workspace navigation and compared the most effective head-based technique against industry standards: \drag and Push--\&--Release. Future research can compare our \head approach to alternatives, such as eye gaze and speech input.

\section{Conclusion}
In this paper, we investigated translating head movements into workspace movements to access off-screen content on wall-sized curved displays. 
We conducted two studies. Results from the first study showed that rate control mapping functions are faster and preferred over zone control functions. 
The polynomial function emerged as most effective, achieving the fastest trial times, the least head rotation, the lowest workload and sickness scores, and the highest user preference. 
To validate these findings in a realistic scenario, our second study compared the polynomial head-based workspace navigation technique with industry-standard controller baselines, drag-and-flick, and joystick-push in a map navigation task.
Results showed that the head-based technique outperforms the baselines in terms of speed, motion efficiency, perceived performance, and user preference, while maintaining high accuracy.
These results offer valuable insights into designing head-based navigation for large, curved displays.

\begin{acks}
This research was funded by NSERC Discovery Grant (\#RGPIN-2019-05211) and the Canada Foundation for Innovation Infrastructure Fund (\#40440).
\end{acks}





\bibliographystyle{ACM-Reference-Format}
\bibliography{main.bib}
\end{document}